\documentstyle[12pt,epsfig]{article}
\newcommand{\eql}[1]{\label{eq:#1}} 
\newcommand{\req}[1]{(\ref{eq:#1})} 
\date{} 
 
\title{ 
     Evaporation of light particles from a hot, deformed and rotating nucleus 
\footnote{Partially supported by the Polish State Committee for Scientific  
          Research  under contract No. 2~P302~052~06}} 
 
\author{                  K. Pomorski                                      \\ 
{\it Department of Theoretical Physics, M. Curie--Sk\l odowska University} \\ 
{\it ul. Radziszewskiego 10, 20-031 Lublin, Poland}                 \\[ 3.0ex]
                     J.~Bartel, J.~Richert                                 \\ 
{\it Laboratoire de Physique Th\'eorique, Universit\'e Louis Pasteur,      } \\
{\it 3, rue de l'Universit\'e, 67084 Strasbourg Cedex , France}              \\
{\it                          and                                         } \\
{\it Centre de Recherches Nucl\'eaires, BP 28, F-67037 Strasbourg C\'edex 2}\\
                                                                     \\[ 3.0ex]
                          K.~Dietrich                                      \\ 
{\it Physik Department,  Technische Universit\"at M\"unchen,              }\\
{\it James Franck Stra{\ss}e, D-85747 Garching, F.R.G.}} 
 
\begin{document} 
 
\maketitle 
 
\vspace{1cm} 
 
\begin{abstract} 
The dependence of the transmission coefficient on the deformation, 
the collective rotation and excitation energy of the compound nucleus 
emitting light particles is introduced in the framework of Wei{\ss}kopf's 
evaporation theory. The competition between fission and particle evaporation
is treated by a~Langevin equation for the fission variable coupled to the 
emission process. Detailed calculations are presented on the decay of 
different Gd and Yb isotopes at an excitation energy of about 250~MeV. 
These calculations demonstrate the importance of the effects of nuclear 
deformation and of the initial spin distribution on the evaporation. 
\end{abstract}

\newpage

\section{Introduction} 
 
The properties of excited nuclei in thermal equilibrium are of 
great physical interest. We are interested in excitation energies below 
some (400--500) MeV, where we may still describe the nucleus as 
a~system of neutrons and protons which interact by effective 
forces. All the information on the physical state of hot nuclei 
is then to be obtained from a~careful study of their decay by 
emission of neutrons, protons, and $\alpha$--particles and their 
decay by fission. The emission of photons will be neglected, as 
we consider energies far above the thresholds for particle 
emission.  
 
The basis of our approach \cite{Str91} is a description of the particle 
emission as a~purely statistical process, as given by the Wei{\ss}kopf 
theory \cite{Wei39}, and the nuclear fission as a~transport process 
\cite{Kra40}. The importance of the non--statistical aspects of the 
fission process in this context was recognized by Grang\'e and 
Weidenm\"uller \cite{Gra79}. Our approach and also the ones of Fr\"obrich, 
Abe and Carjan \cite{Fro92,Til92,Abe90} are based on the premises of 
their work. 
 
Our efforts are dedicated to a stepwise improvement of this 
general theory in order to reach a~quantitatively reliable 
description of the decay of the compound nuclei.  
 
In the present paper, we give a more careful study of how the evaporation
probabilities depend on the deformation of the nucleus, on its excitation 
and on its collection rotation. In the standard form of 
the evaporation theory \cite{Wei39}, the deformation of the decaying 
nucleus enters only through the level densities of the initial 
and final nucleus. We take into account that the transmission 
coefficient also depends on the deformation and, to a~smaller 
degree, on the collective rotation of the nucleus. This will be 
explained in Chapter 2. 
 
Our research is encouraged by the increasing amount of careful experimental 
studies of the evaporation of light particles from excited nuclei and of 
the concomitant decay by fission \cite{Hil92}. The results of our 
calculations are presented and compared with experimental work, especially 
of that of Ref. \cite{Gon90} in Chapter 3. 

Finally, we summarize in Chapter 4 our findings and point out the
direction of our future work.

\vspace{1cm} 
\begin{center}
\section{ Description of Fission Dynamics Including Evaporation }
\end{center}

The fission process of hot and rotating nuclei should be described 
within a statistical model which takes into account the effect of 
energy dissipation due to the coupling to the internal degrees of 
freedom modelled by a friction force which generates diffusion.
The corresponding transport equation of the Fokker-Planck (FPE) type  
was originally proposed for nuclear fission by Kramers  
in Ref. \cite{Kra40} and later on adapted for heavy ions physics in Ref.  
\cite{Hof76} - \cite{Nor79}. 
It is not easy to solve this equation exactly when the number of 
collective coordinates and conjugate momenta is larger than 2.  
An effective but approximate method of solving the FPE based on the moment  
expansion was used in \cite{Ngo77}. Unfortunately the probability  
distributions in the multidimensional space of collective coordinates is  
usually far from a gaussian form and the moment expansion method fails 
\cite{Hof79}. 
It is important to recall here that the set of differential equations which 
one obtains in the moment expansion method for the average coordinates 
and momenta is simply the set of classical equations of motion with friction
as e.g. those used by B\l ocki and co-workers \cite{Blo86}.
A more efficient way of solving the transport equations based on Monte Carlo 
method has been proposed in Ref. \cite{Lam83}. It was also shown in 
\cite{Str91} that the Monte Carlo calculation combined with the local moment 
expansion method transforms the FPE into a set of equations of motion 
containing 
random forces which is equivalent to a set of coupled Langevin equations. 
The Langevin equation has already been used in \cite{Fro88} in order to 
describe heavy-ion collisions and fission \cite{Abe86}. 
This idea is very fruitful and allows us to describe many phenomena occurring 
in heavy-ion physics and fission due to statistical fluctuations 
(see e.g. the review article \cite{Fro92}). 

In Chapter 2 we will present the physical basis of our model
and explain how we couple the emission of light particles to the dynamical 
evolution of the system from the compound nucleus state to the scission point. 
This dynamics is governed in our description by the Langevin equation 
which we will present in Sect. 2.1, before explaining in Sect. 2.2 how 
particle emission is incorporated into this description. In Sect. 2.3
and 2.4 we deal with the initial conditions and the integration of the
Langevin equation.
                                                                    \\[ 0.0ex]

\subsection{ Equations of Motion }

We need to describe the time evolution of the nuclear system from an initial 
state which corresponds to a compound-nucleus at high excitation and angular 
velocity but usually created close to spherical symmetry to a final state in 
which a large amount of excitation and angular momentum will 
have been dissipated. For such a description we use a single collective 
coordinate $q=\rho_{cm}/R_0$ which 
measures the distance between the centers of mass of the two halves of the 
fissioning nucleus in units of the radius $R_0$ of the corresponding 
spherical nucleus. For the time being we restrict our 
description to symmetric fission. The description of asymmetric fission would
necessitate the introduction of an additional collective coordinate describing
the mass asymmetry of the two emerging fission fragments.
If $p$ designates the conjugate momentum associated with the collective 
coordinate $q$ we obtain the following equations 
of motion describing the time evolution of the fissioning nucleus:
\begin{equation}
   \frac{d~q}{d~t} = \frac{p}{M(q)} \;\; ,
\eql{lang1}
\end{equation}
\begin{equation}
   \frac{d~p}{d~t} = {1 \over 2} \biggl( {p \over M(q)}\biggr)^2 \, 
   \frac{d~M}{d~q} - \frac{d~V}{d~q} - {\gamma(q) \over M(q)}\, p + F_L(t) \; .
\eql{lang2}
\end{equation}
                                                                    \\[ 0.5ex]
In these equations $M(q)$ represents the collective mass, $V(q)$ the potential 
and $\gamma(q)$ the friction coefficient \cite{Bar95}. The potential $V$ is 
calculated as the difference between the Helmholtz free energies of the 
deformed and spherical nucleus. The friction 
coefficient $\gamma$ is calculated in the framework of the {\it wall} 
and {\it window} friction model \cite{Blo77,Blo78}. The explicit expressions 
of these quantities, as obtained in the framework of the 
Trentalange-Koonin-Sierk shape parametrization \cite{Tre80}, 
have been presented and discussed in detail in Ref. \cite{Bar95}.
The quantity $F_L(t)$ designates the random Langevin force which couples the 
collective dynamics to the intrinsic degrees of freedom. This Langevin force 
is chosen to be a gaussian random variable with zero mean value. In practice 
eqs. \req{lang1}~-~\req{lang2} are discretized by introducing a sufficiently 
small time step $\tau$. Then the equations take the form
\begin{equation}
   q(t + \tau) - q(t) \approx \frac{p}{M(q)} \tau
\eql{ham1}
\end{equation}
\begin{equation}
   p(t + \tau) - p(t) \approx 
          {1 \over 2} ({p \over M(q)})^2 \, \frac{d~M}{d~q} \, \tau 
          - \frac{d~V}{d~q} \, \tau - {\gamma(q) \over M(q)}\, p \, \tau 
          + \sqrt{D(q)} f_L(\tau) \; . 
\eql{ham2}
\end{equation}
                                                                    \\[ 0.5ex]
Here $f_L(\tau) = \sqrt{\tau} \eta$ where $\eta$ is a gaussian distributed 
random number with $<\! \eta \!> = 0$ and $<\! \eta^2 \!\!> = 2$ and where 
the brackets represent ensemble averages. The diffusion coefficient $D(q)$ 
appearing in eq. \req{ham2} is related to the friction coefficient $\gamma(q)$ 
through the Einstein relation
\begin{equation}
     D(q) = \gamma(q) \; T  \;\; .                                          
\eql{eins}
\end{equation}
This relation holds in linear response theory as a high-temperature 
approximation of the dissipation-fluctuation theorem. 
In the applications presented below the nuclear temperature will always be 
sufficiently high so that eq. \req{eins} is approximately verified. To speak 
at all about a nuclear temperature already supposes that the excitation of 
the nuclear system is shared to equal parts by all of the nucleonic degrees 
of freedom, so that statistical models apply. To simplify the description 
we consider the excited nucleus in a statistical description as member of a 
{\bf grand-canonical ensemble} which can be characterized by a temperature $T$.
For such a description to make sense, we need to 
assume that the time scale which governs the fission dynamics is sufficiently 
long compared to the one which determines the internal equilibration of the 
nuclear excitation among all the nucleons. If this is the case, we can assume 
that the system can be considered as being continuously at equilibrium. 
In this framework the temperature $T$ is simply a measure of the nuclear 
excitation energy $E^*$ and related to the the latter through the usual Fermi 
gas relation $E^* = a(q) \, T^2$ where $a(q)$ is the level density parameter 
of the considered nucleus at a nuclear deformation characterized by $q$. 
The excitation energy itself is determined by the conservation of the total 
energy as will be discussed in section 2.4 below. 
                                                                    \\[-1.5ex]

\subsection{ Particle Emission }

The process of light particle emission from a compound nucleus is governed 
by the emission rate $\Gamma_{\nu}^{\alpha}$ at which a particle of 
type $\nu$ (neutrons, protons and $\alpha$ particles are considered here) 
is emitted at an energy in the range 
$[{e}_{\alpha} - \frac{\Delta {e}_{\alpha}}{2} \, , 
  {e}_{\alpha} + \frac{\Delta {e}_{\alpha}}{2}]$ 
before the compound nucleus eventually undergoes fission. 

Several theoretical approaches have been proposed in order to describe 
the emission from a deformed, highly excited and rotating nucleus 
\cite{Str91,Fro92,Abe90}. The method which we present below is close in 
spirit to the one used in \cite{Str91}, but in our description the widths 
$\Gamma_{\nu}^{\alpha}$ for particle emission will in addition depend on 
the deformation and angular momentum of the compound nucleus \cite{Die95}.

According to Weisskopf's conventional evaporation theory \cite{Wei39} 
the partial decay 
rate $\Gamma_{\nu}^{\alpha \, \beta}(E^*,L)$ for emission of a light particle 
of type $\nu$ with energy ${e}_{\alpha}$ and orbital angular momentum 
$\ell_{\beta}$ from a compound nucleus with excitation energy $E^*$ rotating 
with an angular momentum $L$ can be written as
\begin{equation}
   \Gamma_{\nu}^{\alpha \, \beta}(E^*,L) = \frac{2 S_{\nu} + 1}
              {2 \pi \hbar \rho(E^*,L)} 
              \sum\limits_{L_{R} = |L - \ell_{\beta}|}^{L + \ell_{\beta}} \;\;
   \int\limits_{{e}_{\alpha} - \Delta  {e}_{\alpha}/2}
              ^{{e}_{\alpha} + \Delta  {e}_{\alpha}/2} 
           w_{\nu}({e}, \ell_{\beta}; \chi) \, \rho_R(E^*_R,L_R) \; 
           d \, {e} \;\; ,
\eql{gamma1}
\end{equation}
where
\begin{equation}
   \rho(E^*,L) = (2 L + 1) (\frac{\hbar^2}{2 \cal{J}})^{3/2} \sqrt{a} \;\,
                 \frac{e^{2 \sqrt{a E^*}}}{12 E^{*^{2}}}
\eql{rho}
\end{equation}
is the level density in the emitting nucleus. The level density $\rho_R$ in the
residual (daughter) nucleus with the excitation energy $E_R$ and the
angular momentum $L_R$ is obtained in the same way. Both these quantities 
also depend on the mass and charge number and on the nuclear 
deformation. The quantities $\cal{J}$ and $a$ represent the moment of inertia 
and the level density parameter of the compound nucleus
and $S_{\nu}$ the intrinsic spin of the emitted particle.
$w_{\nu}({e}, \ell; \chi)$ is the transmission coefficient for 
emitting a particle of type $\nu$, with energy ${e}$ and angular 
momentum $\ell$ from the deformed compound 
nucleus. The parameter $\chi$ in the argument list of $w_{\nu}$ stands for 
all quantities not explicitly mentioned here, such as the mass and charge 
number, the nuclear deformation, the direction in space in
which the particle $\nu$ is 
emitted. Proceeding in this way would, however, leads to hardly tractable 
numerical problems, since in the Langevin formalism which we endeavour here, 
we need to follow the dynamics of the fissioning nucleus plus evaporation for 
a very large number of trajectories (of the order of 10$^6$). We therefore 
use a simplified procedure which introduces a transmission coefficient 
$\bar{w}_{\nu}({e}, \ell; \chi)$ obtained by a double averaging over the 
different emission directions and over the whole surface of the deformed 
compound nucleus. A detailed description of how  
$\bar{w}_{\nu}({e}, \ell; \chi)$ is calculated in the framework of the 
Hill-Wheeler approximation \cite{Hil53} and how the averaging procedure is 
carried out is given in Appendix A.
With this averaged transmission coefficient $\bar{w}$ the width 
$\Gamma_{\nu}^{\alpha}$ for emission of a particle of type $\nu$ and energy 
${e}_{\alpha}$ reads:
\begin{equation}
   \Gamma_{\nu}^{\alpha}(E^*,L) = \frac{2 S_{\nu} + 1}
              {2 \pi \hbar \tilde{\rho}(E^*)} 
   \int\limits_{{e}_{\alpha} - \Delta  {e}_{\alpha}/2}
              ^{{e}_{\alpha} + \Delta  {e}_{\alpha}/2} 
        w^{eff}_{\nu}({e}; \chi) \, \tilde{\rho}_R(E^*_R) \; 
           d \, {e} \;\; ,
\eql{gamma2}
\end{equation}
where 
$$   \tilde{\rho}(E^*) = \frac{\rho(E^*,L)}{2 L + 1}            \nonumber    $$
is the angular momentum independent part of the density \req{rho} and a similar 
relation holds for $\tilde{\rho}_R$. The effective transmission coefficient 
in \req{gamma2} is obtained by performing a summation over all allowed angular 
momenta of the emitted particle and those of the daughter nucleus:
\begin{equation}
w^{eff}_\nu(e; \chi) = \frac{1}{2L + 1} \;
        \sum\limits_{\ell_\beta = 0}^{\ell_{max}} \;
        \sum\limits_{L_{R} = |L - \ell_{\beta}|}^{L + \ell_{\beta}} \;
        (2L_R + 1) \bar{w}_\nu(e,\ell_\beta ; \chi) \;\;\; ,
\eql{weff}
\end{equation}
here $\ell_{max}$ is the maximal angular momentum available for the particle
having the energy $e$.
As already mentioned the 
emission width $\Gamma_{\nu}^{\alpha}$ also depends on the mass and charge 
number A and Z as well as on the deformation of the compound nucleus. 

Once the emission widths $\Gamma_{\nu}^{\alpha}$ known, one can establish 
the emission algorithm which decides at each time step $[t, t + \tau]$ along 
each of the trajectories if a particle is emitted from the compound 
nucleus. It is the value of the emission width 
$\Gamma_{\nu}^{\alpha}$ which ultimately have to decide which of the light 
particles is emitted and at which energy. Since $\Gamma_{\nu}^{\alpha}$ 
represents the rate at which a particle of type $\nu$ and energy ${e}_{\alpha}$ 
is emitted, the total emission rate for a particle of given type, 
irrespectively of the energy at which it is emitted, is given by 
\begin{equation}
   \Gamma_{\nu}(E^*,L) = \frac{2 S_{\nu} + 1}
              {2 \pi \hbar \tilde{\rho}(E^*)} \int\limits_{0} ^{e_{max}}
        w^{eff}_{\nu}({e}; \chi) \, \tilde{\rho}_R(E^*_R) \, d \, {e} \, 
        = \, \sum\limits_{\alpha = 1}^{n}\Gamma_{\nu}^\alpha (E^*,L) \,\, ,
\eql{gamma3}
\end{equation}
where we have replaced the upper integration limit by some large enough 
constant ${e}_{max}$ (which will, a priori, be different for the different 
particles) and at which the probabilities for particle emission will 
essentially have vanished. $n$ is the number of energy bins of width 
$\Delta{e}_\alpha$ in the interval $[0, e_{max}]$. 
The emission rate $\Gamma$ for emission of any kind of particles is
the sum of the $\Gamma_{\nu}$:
\begin{equation}
   \Gamma = \Gamma_{n} + \Gamma_{p} + \Gamma_{\alpha} \;\; .
\eql{gamma4}
\end{equation}
First of all we need to decide whether a particle is emitted at all in 
the given time interval [t, t + $\tau$]. The probability for emitting 
any particle is given, for a small enough time step $\tau$, by
\begin{equation}
   P(\tau) = 1 - e^{-\Gamma \, \tau} \approx \Gamma \, \tau \;\; .
\eql{prob}
\end{equation}
One then draws a random number $\eta_1$ in the interval $[0,1]$. If 
$\eta_1 \!\! < \!\! P(\tau)$ a light particle is emitted. If the time step 
$\tau$ 
is chosen sufficiently small, the probability for emitting a particle will be 
small. In this way we guarantee that in a time interval at most one particle
is emitted and we avoid to consider the emission of more than one particle in 
each time interval. 

In the case that a particle is emitted one needs to decide 
next of which type this particle is. To this purpose one draws a second random 
number $\eta_2$ in the interval $[0,1]$ and determines the localisation of 
$\eta_2$ with respect to the covering of this interval by the three bins 
$\Gamma_n/\Gamma$, $\Gamma_p/\Gamma$, and $\Gamma_{\alpha}/\Gamma$. Depending 
on the bin in which $\eta_2$ is located, a neutron, or a proton or an $\alpha$ 
particle is emitted. 

We still have to determine the energy with which this particle is emitted,
and we accomplish this in the following way. We introduce the quantity
\begin{equation}
   \Pi_{\nu}({e}_{\alpha}) = \frac{1}{\Gamma_{\nu}} \,\left\{ 
       \frac{2 S_{\nu} + 1}{2 \pi \hbar \tilde{\rho}(E^*)}
       \int\limits_{0} ^{e_\alpha} 
       w^{eff}_\nu (e; \chi) \, \tilde{\rho}_R(E^*_R) \, d \, {e}\right\} \,\, ,
\eql{pi}
\end{equation}
which represents the probability that a particle of type $\nu$ is emitted with 
an energy smaller than ${e}_{\alpha}$. The quantity 
covers the interval $[0,1]$. Subdividing the interval $[0,1]$ in a certain 
number of equal bins, one decides with which energy the light particle is 
emitted by drawing a third random number $\eta_3$ in the interval $[0,1]$. 
Inverting the function in eq. \req{pi} $\Pi^{-1}_{\nu}(\eta_3)$ will fall 
in one energy bin 
\begin{equation}
    \Pi^{-1}_{\nu}(\eta_3) \in [{e}_{\alpha} - \frac{\Delta {e}_{\alpha}}{2}, 
                                {e}_{\alpha} + \frac{\Delta {e}_{\alpha}}{2}]
\eql{ealpha}
\end{equation}
of the total interval $[0,{e}_{max}]$ and thus decide on the energy with 
which the particle is emitted.
                                                                    \\[-1.5ex]

\subsection{ Initial Conditions }

The description of the fission process including particle emission starts, 
in principle, from an initial state corresponding to a spherical or deformed 
compound nucleus whose shape is characterized by the collective coordinate 
$q_0$, the corresponding conjugate initial momentum $p_0$, the excitation 
energy $E_0^*$ (with corresponding temperature $T_0 = \sqrt{E_0^*/a(q_0)}$) and
the initial total angular momentum $L_0$. In practice the excitation energy 
$E^*$ may be known approximately from the experiment. This is, however, not 
the case for $q_0$, $p_0$ and the angular momentum $L_0$.
In the calculations presented and discussed in Sect. 4 we 
choose $q_0$ corresponding to a spherical nucleus and draw the conjugate 
momentum $p_0$ for each trajectory from a normalized gaussian distribution
\begin{equation}
P(p_0) = \frac{1}{\sqrt{2 \pi m T_0}} \exp[-\frac{p_0^2}{2mT_0}]\,\, .
\eql{Pini}
\end{equation}
Here $m$ is the collective inertia at the deformation $q_0$.
It is in principle experimentally
possible to determine the distribution of initial angular momenta for instance 
by measuring the $\gamma$ multiplicities \cite{Kuh89}. As these distributions 
are rarely available, we chose $L_0$ to have a fixed value but we shall 
discuss below the dependence of the final results on the choice of $L_0$. 
                                                                    \\[-1.5ex]

\subsection{ Integration of the Langevin Equation }

Once the initial conditions are fixed one can integrate the system of 
equations of motion (1-2) using their finite difference version (3-4). At 
each time step $[t,t + \tau]$ one draws a random number $\eta$ from a gaussian 
distribution which defines the fluctuating force in eq. (4) and thus generates 
a trajectory in the variable $q$. Simultaneously one asks, as just explained, 
whether a light particle is emitted and decides, eventually, of its type $\nu$ 
and energy ${e}_{\alpha}$.

At the beginning of the whole process the system has a fixed total energy 
E$_{tot}$ which is the sum of the collective kinetic and potential energy 
$$   E_{coll} = \frac{p^2}{2 \, M(q)} + V(q)             \nonumber        \\ $$
and a collective rotation energy 
$$   E_{rot} = \frac{L^2}{2 \, \cal J}                   \nonumber        \\ $$
corresponding to a rigid deformed rotator of moment of inertia $\cal J$
\cite{Ben94,Cha94}. The excitation energy E$^*$ at time 
$(t + \tau)$ and thereby the temperature of the nucleus is redetermined at 
each time step through the conservation of the total energy
\begin{equation}
   E_{tot} = E_{coll}^{(i)} + E_{rot}^{(i)} + E^*_i 
           = E_{coll}^{(f)} + E_{rot}^{(f)} + E^*_f  
                            + {e}_{\alpha} + B_{\nu} + E_{recoil} \;\; ,
\eql{etot}
\end{equation}
where the indices i or f refer to the initial (time $t$) or final state (time
$t + \tau$). $B_{\nu}$ is the binding energy of the emitted particle (which 
is zero unless an $\alpha$ particle is emitted), ${e}_{\alpha}$ is its 
kinetic energy after emission and $E_{recoil}$ is the recoil energy of the 
nucleus due to the emission process. These three quantities enter 
the equation only when a light particle is emitted in the time interval 
$[t,t + \tau]$. Since the recoil energy is in any case very small, it
will be neglected. Additionally it is assumed that the deformation as well as 
the angular frequency of the nucleus is not changed due to the particle 
emission. 

It is evident that each emission of a light particle carries away excitation 
energy and angular momentum and thereby increases the height of the fission 
barrier of the residual nucleus which, in turn, renders the fission event 
less and less probable.

For each choice of the initial conditions one generates in this way a separate 
trajectory which is followed through in time. Emitted particles and their 
energies are registered. Each trajectory can either lead to fission if it 
overcomes the fission barrier and continues on to the scission point, or can 
end up as a rather cold compound nucleus if too much energy has been lost to 
make the crossing of the fission barrier possible. Sampling the total number 
of trajectories $N_{fiss}$ which have led to fission defines the fission cross 
section as $\sigma_{fiss} = N_{fiss}/N$ with $N$ equal to the total number 
of trajectories.

Let us finally mention that it is important to take into account the fact that 
the $\alpha$ is a composite particle and that its emission therefore 
presupposes its existence in the nucleus prior to emission. This is in 
principle quantified by the introduction of a preformation factor whose value 
can in principle be determined \cite{Die96}. Since this quantity is, 
however, 
very poorly known, we fix it in the following by multiplying the emission 
width $\Gamma_{\alpha}$ by a factor $f_{\alpha}$ ($0 \leq f_{\alpha} \leq 1$).
The choice of $f_\alpha$ will be discussed in the following.

\vspace{1cm} 

\section{Results} 
 
We study the decay of compound nuclei of several isotopes of $_{64}$Gd 
and of $_{70}$Yb at various excitation energies ranging from 150 MeV to 
300 MeV. We selected these nuclei because a~careful experimental 
investigation of their decays is available [7] and because they are 
known to exhibit large ground state deformations. Consequently, it is 
of special interest to investigate the influence of deformation on the 
emission of $n$, $p$ and $\alpha$ particles from excited states of these 
nuclei.  
 
The results concern the following physical aspects: 
\begin{itemize} 
\item[~~i] The dependence of the emission probabilities of neutron, 
protons, and $\alpha$--particles on the deformation, the excitation energy
and the angular momentum of the emitting 
nucleus (Figs. 1 to 9). 
\item[~ii] The dependence of the emission probabilities $\Pi_\nu({e})$, 
eq. \req{pi}, of a particle of given type $\nu$ on the kinetic energy ${e}$ 
(Figs. 10 to 15). 
\item[iii] The dependence, at different temperatures $T$, of the number of 
fission events and light-particle multiplicities on the time t 
(Figs. 16 to 18). 
Although these functions are not measurable, they are important for the 
physical understanding of the decay, especially the transient time 
phenomenon [4] and its dependence on the temperature. 
\item[~iv] The spectral distribution of the emitted particles (Figs. 19 to 21).
\item[~~v] The dependence of the multiplicities, fission cross section and 
barriers on the initial angular momentum of the nucleus (Figs. 22 to 25).
\item[~vi] The influence of the friction forces on the multiplicities of the
emitted particles (Fig. 26).
\end{itemize} 
We will now give a detailed description of these points. 
 
\subsection{Dependence of the emission probabilities on the nuclear 
            deformation}  
 
In Figs.\ 1--3, the emission width $\Gamma_\nu$, eq. \req{gamma3}, 
for $n$, $p$, and $\alpha$--particles is shown as a~function of 
the deformation $\rho_{cm}/R_0$ of the emitting source. The initial ensemble of 
decaying nuclei consists of $^{160}_{70}$Yb$_{90}$ nuclei at an 
excitation energies of 50 MeV, 150 MeV and 250 MeV respectively, and with 
a~rotational angular momentum $L = 40\hbar$. All the emission rates
are seen to grow as a~function of increasing
deformation. This trend can be easily understood as for increasing deformation 
the transmission occurs through a~larger surface. This effect has already been 
observed for all three particles \cite{Bla80}. 
We notice, however, that the emission width for $\alpha$--particles increases 
more steeply than the one for $n$ and $p$ for the two lower excitation 
energies $E^\star$ = 50 MeV (Fig. 1) and $E^\star$ = 150 MeV (Fig. 2). 
This is due to the fact that, as the nucleus is elongated, the barrier 
height for charged particles is reduced in the section of the surface 
which is farther away from the nuclear center and increased 
in the section which are closer to the center. The section where
the barrier is diminished is larger than the section where it is 
increased. Consequently, the emission rate for charged particles 
increases faster than the one for neutrons.

In Figs. 4, 5, and 6 the deformation dependences of the emission
rates for $n$, $p$, and $\alpha$ particles are shown for different 
values of the rotational angular momentum, varying between $L = 
0\hbar$ and 60~$\hbar$. For $n$ and $p$ the influence of the rotation 
on the emission rate is seen to be negligible, whereas the emission 
rate of $\alpha$ particles grows by 10\%--20\% as the rotational
angular momentum increases from 0~$\hbar$ to 60~$\hbar$. 
It is clear that the centrifugal force, which helps to overcome
the barrier, is largest for the $\alpha$--particle. Furthermore,
the centrifugal effects matter the more the larger the barrier.
The barrier for the $\alpha$--particle is larger than the one
for $n$ and $p$. 

Let us point out that the rotational angular momentum of the
nucleus has a very noticeable influence on the height of the
fission barrier which decreases as a function of increasing
angular momentum. Thus, at high angular momentum, nuclear
fission can compete more effectively with evaporation. 

In Figs. 7--9, the deformation dependent emission width for
$n$, $p$ and $\alpha$ is shown for two different isotopes of both 
$_{64}$Gd and $_{70}$Yb. The results can be easily understood: 
for given proton number, the emission width $\Gamma_n$ for
neutrons is the larger the larger the neutron number (see Fig. 7). On the
other hand, for given neutron number, the emission width
$\Gamma_p$ for protons is the larger the larger the proton
number (see Fig. 8). At given proton number, the emission width
$\Gamma_\alpha$, for $\alpha$--particles decreases with increasing
neutron surplus whereas, at given neutron number,
$\Gamma_\alpha$ grows with increasing proton number (see
Fig. 9). This is a simple consequence of the fact that the
$\alpha$--particle contains an equal number of neutrons and
protons.

\subsection{Behaviour of the probability $\Pi_{\nu}(e)$} 

An interesting quantity is the probability
$\Pi_\nu({e})$ to emit a particle of type $\nu$ with
an energy smaller than ${e}$. Its definition in
terms of the emission rate $\Gamma_\nu({e})$ is given
in eq. \req{pi}. The emission width $\Gamma_{\nu}$ depends also
on the shape of the emitting nucleus, and so does the
integrated probability $\Pi_\nu({e})$.

In Figs. 10 to 12, the probability
$\Pi_\nu({e})$ is shown for the emission of
neutrons, protons, and $\alpha$--particles. In each of the
figures, $\Pi_\nu({e})$ is plotted separately for the
case that the emitting nucleus has a spherical shape, the shape
corresponding to the saddle and the scission point.
In all cases the fissioning nucleus is
$^{160}$Yb at an initial excitation energy $E^{\star} = 250~MeV$
and an initial angular momentum $L = 40 \hbar$. All curves show
a monotonous increase of $\Pi_\nu$ as a function of the energy
${e}$ starting from a minimal energy
${e}$ which is zero for neutrons, but finite for
protons and $\alpha$--particles due to the acceleration of the
charged particles in the Coulomb field of the residual nucleus. 
For neutron emission, the dependence of $\Pi_n({e})$ on the
deformation of the emitting nucleus is very small, whereas for 
protons and $\alpha$--particles the curves $\Pi_\nu({e})$ are shifted 
to somewhat higher energies for the spherical source. The
physical interpretation is that the Coulomb barrier is larger
for the spherical nucleus than for the deformed nucleus and
consequently the threshold for proton or $\alpha$--emission is
at a higher energy than for neutrons. We note that
the scission shape for a nucleus like $^{160}$Yb consists
essentially of two tangent, slightly deformed fragments of half
the charge.

In Figs. 13 to 15 the probability $\Pi_\nu({e})$ is shown for neutrons, 
protons, and $\alpha$--particle emission for 3 different excitation
energies of the initial compound nucleus $^{160}_{70}$Yb. The shape of 
the emitting nucleus is chosen here as corresponding to the saddle point
and its initial angular momentum is fixed at $L = 40~\hbar$.
It is seen that the rise of the probability $\Pi_\nu$ as a
function of ${e}$ is the steeper the smaller the excitation energy. 
Such a behaviour can be easily understood since the
range of energies of the emitted particles must rise with
excitation energy. We have also investigated the dependence of
the functions $\Pi_\nu({e})$ on the neutron and
proton numbers of the initial ensemble of compound nuclei,
comparing the emission from the 4 nuclei $^{144}_{64}$Gd,
$^{154}_{64}$Gd, $^{150}_{70}$Yb, and $^{160}_{70}$Yb, in each
case for the same excitation energy (150 MeV), the same angular
momentum (40 $\hbar$) and the same shape (saddle point). The
results are indistinguishable for the emission of neutrons and
only very slightly dependent on the nature of the emitting
nucleus for the emission of protons and $\alpha$--particles.

\subsection{Time dependence of the multiplicities and the number
            of fission events }

The dependence of the number of decays of a given type on the
time which elapses starting with the formation of the compound
nucleus is unfortunately not measurable. Nevertheless, we think
that it is interesting to exhibit this dependence for a few
cases since it enables us to gain insight into the dynamical mechanism.

In Fig. 16 we show the number of fission
events as a function of the time t on a logarithmic scale. This result was 
obtained with the light-particle evaporation channels turned off. The initial
compound nucleus is $^{160}_{70}$Yb with an initial angular
momentum $L = 40~\hbar$. The 3 curves in Fig. 16 correspond to 3
different initial temperatures resulting in 3 different
initial fission barrier heights $U_B$. It is seen that the transient
time increases with decreasing excitation energy, as one
expects. It should be noted that the functions $N_{\rm fiss}(t)$
obtained for $T = 4$~MeV look quite similar to those for $T = 5$~MeV and 
are just shifted along the $log(t)$ axis. Furthermore, if $t_0$ is the 
time where half of the final number $N_{\rm fiss}(\infty)$ of fission
events have occurred
\begin{equation} 
  N_{\rm fiss}(t_0) = {1 \over 2} N_{\rm fiss}(\infty)   \nonumber
\end{equation}
we may approximate $N_{\rm fiss}$ for times close to $t_0$ by a
linear function of $log(t)$
\begin{equation}
 N_{\rm fiss}(t) \approx N_{\rm fiss}(t_0) + \kappa_0 \, log{t\over t_0}
\eql{3.3}
\end{equation}
As one infers from Fig. 16, the dependence on the initial
temperature is mainly contained in the quantity $t_0$ whereas the
factor $\kappa_0$ in \req{3.3} is almost the
same for $T = 5$~MeV and $T = 4$~MeV. The approximate time
dependence \req{3.3} which is valid during part of the transient
time interval is seen to be totally different from the one in
the Kramers regime  which is valid in the cases when the fission
barrier is much higher than the temperature of the fissioning nucleus.

In Fig. 17 we show the multiplicity (number of emitted particles in
coincidence with fission and per compound nucleus) for the
emission of neutrons, protons and $\alpha$--particles as a
function of time for the decay of a compound nucleus
$^{160}_{70}$Yb at an initial excitation energy of 293 MeV and
an initial angular momentum of 45 $\hbar$. It is seen that the
emission of $p$ and $\alpha$--particles ceases already after some 
3$\cdot$10$^{-21}$~sec whereas the number of emitted neutrons still increases. 
Again this is easily understood as a
result of the different thresholds for charged and uncharged
particles. 

In Fig. 18 the fraction ${N_{\rm fiss}/N}$ of nuclei undergoing fission
is shown for the initial compound nucleus $^{160}_{70}$Yb as a function
of time. Now, contrary to the results presented in Fig. 16, the
emission of light particles is taken into account.  The initial
excitation energy of the nucleus is $E^{\star} = 293$~MeV and the
initial angular momentum $L = 45~\hbar$ is assumed. The well--known
Kramers approximation holds whenever there is an approximately constant
current across the fission barrier which implies a linear dependence of
the number of fission events upon time. It is seen from the figure that
there is in fact no clearly distinguished section with a linear
time-dependence. At best one could replace the function by a straight
line in the interval between 8 and 12$\cdot$10$^{-21}$~sec. The
``transient time'', i.e. the time needed for the build--up of an
approximately constant fission current is seen to be
$\sim$~8$\cdot$10$^{-21}$~sec. This is the main part of the time
available for fission events. It is thus clear that the use of the
Kramers approximation or of the still simpler statistical transition
state hypothesis for the whole time interval would lead to wrong
results.

Furthermore the very concept of a constant current across the fission
barrier, which is the pre-requisite of the Kramers approximation could
loose its validity because of the existence of light particle emission
which implies that the excitation energy and the angular momentum
changes for those compound nuclei which emit a light particle prior to
fission.

\subsection{Spectral distribution of the emitted particles}

In Fig. 19, we present the probability per energy unit
of neutron, proton, and $\alpha$--particle emission as a
function of the energy of the emitted particle.  This figure is made on the 
basis of $10^6$ trajectories. The solid
curves show the spectral distribution in coincidence with
fission, i.e. for particles emitted from nuclei which
subsequently undergo fission. The dashed curves represent the
spectral distribution in anti--coincidence with fission, i.e.
for particles emitted from nuclei which subsequently end up as
evaporation residues. The initial compound nucleus is
$^{160}_{70}$Yb at 293 MeV excitation energy and an angular
momentum of 52 $\hbar$. The curves for different types of emitted
particles are displaced against each other because charged
particles gain energy in the Coulomb field of the residual
nucleus. Furthermore, the thresholds for emission of $n$, $p$,
and $\alpha$ are in general different.

In addition, for $\alpha$--particles, the spectral distribution
for particles emitted in coincidence with fission is slightly shifted 
towards smaller energies as compared to the distribution obtained when 
measured in anti-coincidence with fission. This is
probably due to the fact that charged particles are
preferentially emitted from the pole tips around the long half
axis. The larger deformation then implies a smaller gain of
kinetic energy from the repulsive Coulomb field.

In Figs. 20 and 21, the normalized yield for neutron emission is shown 
as a function of the nuclear deformation in the fission and evaporation 
channels. The reader should notice that for 
nuclei which undergo fission, the emission of neutrons takes
place, on the average, at a larger deformation than for the
nuclei which end up as evaporation residues. This is due to the
fact that the distribution of the fissioning nuclei moves
towards the saddle point in the deformation landscape. This effect 
should give rise to an experimentally observed 
anisotropy in the angular distribution of prefission
particles different from the one observed in the angular distribution of
particles emitted by the evaporation residua. The distribution for 
the case when the emission of protons and alpha particles is inhibited 
is shown in Fig. 20 by the short--dashed line. It is 
seen that this distribution is very close to the one obtained when the
emission of all three kinds of particles is allowed.
The initial angular momentum is different for the two Figs. 20 and 21. It
is seen that the maximum of the distribution of tne neutrons emitted by
the evaporation residua is shifted with increasing 
angular momentum towards larger deformations.
The difference in the average deformation of the both distributions
should be directly reflected in the difference in the anisotropy
of the angular distributions of prefission neutrons and those
emitted by the residua.

\subsection{Dependence of the multiplicities on the initial angular
            momentum of the compound nucleus}

The multiplicity of  prefission neutrons, protons and alpha particles 
is plotted in Figs. 22--23 as a function of the initial angular momentum 
$L$. The initial compound nucleus is $^{160}$Yb with 
the initial excitation energy E$^*$=251 MeV in Fig. 22 and 293 MeV in Fig. 23.
The corresponding fission rates are plotted in Fig. 24.
It is seen in Figs. 22 -- 23 that the neutron multiplicity decreases 
significantly with growing $L$ while that for protons and alphas is much less
affected. This is due to the fact that for large angular momenta the fission
barriers $U_B$ become small (see Fig. 25) and it takes a shorter time to reach 
the scission configuration. Consequently less neutrons are emitted on 
the average when the barrier is small. Alpha particles and protons are 
mostly emitted in the initial stage when the excitation energy of the nucleus 
is large, so that their multiplicities depend more weakly on the initial $L$.
The effective fission rate (including particle emission) changes by 2 orders 
of magnitude (see Fig. 24) when one increases the value of the initial angular
momentum by 12$\hbar$ units up to its maximal value which corresponds to 
$U_B$=0. Fig. 23 shows that this also holds true for the higher excitation 
energy of $E^*=$293 MeV.

This result indicates that a more precise knowledge of the initial spin 
distribution in compound nuclei is necessary in order to compare to the 
experiment. One has to stress here that large $L$ values contribute most 
to the experimental value of the multiplicities as already seen in Fig. 24. 
It therefore is certainly not correct to describe the experiment with a 
theoretical calculation using only the average value of the initial spin. 
As the fission barrier heights decrease strongly with increasing angular 
momentum as seen in Fig. 25, the higher L components which would lead to 
fission will have to be mocked up by an overestimation of the role of 
the fluctuating forces. 

Assuming an equal population of the initial angular momenta, which is, 
as just mentioned, not the best 
approximation, we have estimated the average of the $n$ , $p$ and
$\alpha$-multiplicities. The resulting values are compared in Table 1 with
the experimental data taken from Ref. \cite{Gon90}.  All parameters of the 
model are those given in Sec. 2.1 and in Ref. \cite{Bar95}.
We only have to assume the preformation factor $f_\alpha = 0.2$ (see 
discussion in Sec. 2.4) in order to reproduce the experimental number of alpha 
particles. This goes into the right direction since our calculations show 
that $\alpha$ particle emission is strongly enhanced by rotation and 
deformation effects.

\subsection{Influence of the friction forces on multiplicities of
            prescission particles}
            
The role of the friction forces in fission of high excited compound nuclei is
widely known.  It was shown already in Ref.  \cite{Gra79} that the nuclear
viscosity influence significantly the prescission neutron multiplicities.  This
effect was discussed later in almost all papers dealing with the problem.  It
was even assumed that the neutron multiplicities could be an indirect measure of
the nuclear friction (see e.g.  Ref.  \cite{Hil92}).

The dependence of $n$ , $p$  and $\alpha$-multiplicities on the
magnitude of the friction coefficient connected with the elongation of
the nucleus is plotted in Fig. 26 for $^{160}$Yb at $E^*$=293 MeV and
$L$=50$\hbar$.  The strength of the friction coefficient is varied in
the interval (0.02, 2) in units of its value given by the wall formula
\cite{Blo86}. The neutron multiplicity for small values of the friction
coefficient is about 30\% smaller than obtained with the wall friction.
As the transient time increases as a function of the friction, the
increase of the multiplicity of neutrons emitted prior to fission is easly 
understood. Again the effect upon the emission of $p$ and $\alpha$-particles 
is small, because these particles are emitted in a relatively
short time interval after the formation of the compound nucleus which
is in our case small compared to the transient time. One has to notice
that the results for very weak friction depend significantly both on
the width of the initial distribution of the momentum conjugate with
the fission coordinate (eq. \req{Pini}) and on the initial angular
momentum. For larger values of the friction strength, close to
$\gamma_{wall}$, the influence of the initial distribution of the
momentum $p_0$ on the neutron multiplicity is weak (see also the
discussion in Ref. \cite{Str91}).

\section{Summary and discussion}

Our results demonstrate the importance of nuclear deformation on 
the evaporation of light particles from strongly excited 
nuclei. This dependence on the deformation plays an important
role also for the competition between fission and particle
emission and might modify the limits which were determined for the 
nuclear friction force \cite{Str91} from the experimental data 
[6] on evaporation and fission. The dependence of the evaporation 
width on the rotational angular momentum was found to be 
negligible for the emission of $n$ and $p$ and rather small for 
the emission of $\alpha$--particles. Let us, 
however, remind that the fission probability 
depends very sensitively on the angular momentum, as we
have already emphasized in Ref. \cite{Str91}. Due to this strong dependence
of the fission probability on the initial angular momentum, it
becomes very important to get precise information on the angular
momentum distribution of the initial ensemble of compound
nuclei. The outcome of the competition between light particle
emission and fission may depend strongly on the initial angular momenta.
Consequently, we have to try to obtain theoretical information on the 
angular momenta of the initial nuclei by treating the fission process 
dynamically \cite{Prz94,Pom94}.

Agreement between calculated and measured emission probabilities
for $\alpha$--particles can only be obtained, if an empirical
preformation probability for $\alpha$--particles of about 0.2 is
assumed. One of the most important further improvements of the
theory will be to evaluate this preformation factor within the
temperature--dependent Thomas--Fermi approximation which
underlies our theory \cite{Die96}.

Another aspect which we intend to investigate is the angular 
dependence of emitted light particles when the
angular momenta of the initial compound nuclei are aligned or the nuclei are 
polarized. In these cases it is conceivable that the angular dependence of the 
emitted particles shows a more pronounced dependence on the
rotational angular momentum than the integrated yields we
calculate in the present paper. One might also hope that the angular
distribution of emitted particles depends on the deformation of
the source nuclei sufficiently sensitively so as to determine
the deformation from such measurements. Of course, experimental
data on the angular distribution of emitted neutrons, protons,
and $\alpha$--particles from aligned rotating deformed nuclei
would be of great interest for these studies.

\vspace{20mm} 
 
\noindent 
{\bf Acknowledgments} 
 
One of us (K.P.) acknowledges gratefully the support during a four week 
stay in Garching by the Technical University Munich and support  
by the IN2P3 during a one month stay at C.R.N. in Strasbourg. 
J. B., K. D., and J. R. are very grateful for
the kind hospitality extended to them at the UMCS. K.~D.
acknowledges financial support from the BMFT and also from the
DAAD for the travel funds to Lublin.

\vfill\eject

{\large\bf Appendix: Evaluation of the transmission coefficients $w_\nu$}
                                                                     \\[ 1.0ex]

The explicit calculation of the transmission rates $\Gamma_{\nu}^{\alpha}$ 
given by eq. \req{gamma2} requires the knowledge of the average transmission 
coefficient $\bar{w}_{\nu}({e},\ell ; \chi)$ for the emission of light particles 
from the nucleus into the continuum. In principle, a rigorous treatment 
would require the knowledge of $w_\nu({e},\ell ; \chi)$ which appears 
in eq. \req{gamma1} and should be calculated by averaging the transmission 
coefficients corresponding to the transmission of a particle emitted in
a given direction from each 
point of the surface of the deformed, excited and rotating nucleus. 
This can be done in principle but is hardly possible in practice where 
these calculations must be carried out at each time step and over a 
very large ensemble of trajectories. We introduce the following 
simplified procedure which follows four steps:
\begin{itemize}
\item
We first calculate the transmission coefficients at three selected points 
(1,2,3) at which the main body-fixed axes (x,y,z) respectively cross the 
nuclear surface (see Fig. A1). 
The coordinate system is chosen such that the rotation axis coincides with 
the x-axis and the z-axis is the symmetry axis.

For all possible values of the angular momentum of the emitted particle
$$   \ell = | \vec{r}_i \times \vec{p}_i | = r_i \; p_{i_{\parallel}}\,\, ,  
\eqno(A1)
$$
where $p_{i_{\parallel}}$ is the component of the vector $\vec{p}$ parallel to 
the nuclear surface, and for each of the points (i=1,2,3) $w$ is calculated 
using the Hill-Wheeler WKB expression \cite{Hil53}
$$
   w_\nu({e}, \ell, \ell_x; i) 
     = \biggl( 1 + exp[-\frac{2\pi (E_B-e)}{\hbar \omega_i}] \biggr)^{-1} \;\; .
\eqno(A2)
$$
Here $e$ is the single-particle energy and the barrier height $E_B$ corresponds
to the maximum of the potential $V_{tot}$ in the direction $r_{i_{\bot}}$ 
perpendicular 
to the surface. The potential in which the particle moves is given by:
$$
V_{tot} =  V_{nucl} + V_{cent} + V_{Coul} - \omega \ell_x 
\eqno(A3)
$$
i.e. by the sum of nuclear \cite{Str91}, centrifugal and Coulomb (in the 
case of charged particles) single-particle potentials. The quantity 
$\hbar \omega_i$ appearing in (A2) is given by 
$$ 
\hbar \omega_i = 
     \hbar \sqrt{\frac{d^2 V_{tot}(i) / d r_{i_{\bot}}^2}{m_\nu}}\,\, .
\eqno(A4)
$$
\item
Once $w_\nu({e}, \ell, \ell_x; i)$ is fixed at the three points (i=1,2,3) 
one averages in a second step over all possible values of the x-component 
of $\ell$ which is the component along the rotation axis. 
Such a procedure seems to be reasonable once the angular momentum of 
the emitted particles is not measured experimentally.

Through this average procedure one obtains
$$
\bar{w}_\nu({e},\ell; i) = \frac{\sum\limits_{\ell_x = -\ell}^{\ell} 
       w_\nu({e}, \ell, \ell_x; i)}{2 \ell + 1} \;\;\; ,\;\; i=1,2,3 \; . 
\eqno(A5)
$$
\item
The transmission coefficients $\bar{w}_\nu({e},\ell; i)$ being determined 
at the three selected points (i=1,2,3) one carries out in a third step 
an interpolation which allows to calculate the transmission coefficients 
to any point $(\theta, \varphi)$ of the nuclear surface 
$$     \bar{w}_\nu({e},\ell;\; \theta, \varphi) = sin^2 \, \theta 
       \cdot (\bar{w}_1 cos^2 \, \varphi + \bar{w}_2 sin^2 \, \varphi) 
                +  \bar{w}_3 cos^2 \, \theta \,\, .
\eqno(A6)
$$
This interpolation of quadrupole type ensures that the transmission coefficient 
in the directions along the main axes are the same as those for emission in the 
opposite direction. 
\item
Finally one averages the coefficients over the whole nuclear surface: 
$$
\bar{w}_\nu({e},\ell; \chi) = 
      \frac{\int\limits_S \bar{w}_\nu({e},\ell;\; \theta, \varphi)\, d\sigma}
           {\int\limits_S\, d\sigma} \,\, .
\eqno(A7)
$$
\end{itemize}
The procedure described above incorporates in an approximate way the effects 
of nuclear deformation and rotation on the transmission coefficients.

\vfill\eject
\parindent=0cm
{\bf\large Table caption}
\begin{enumerate}
\item Results of prescission neutron, proton and alpha multiplicity 
      model calculations for $^{160}$Yb at excitation energies 251 MeV and 
      293 MeV compared with the experimental data taken from 
      Ref. \cite{Gon90} .The theoretical results are averaged over all 
      initial angular momenta with a weight proportional to the 
      corresponding fission rates.
\end{enumerate}

\vspace{2cm}
{\bf\large Figures captions}
\begin{enumerate}
\item Emission widths for neutrons (n), protons (p) and alpha 
         particles ($\alpha$) emitted from the hot, rotating compound nucleus 
         $^{160}$Yb (E$^*$=50 MeV, L=40$\hbar$) as a function of the 
         elongation.
\item The same as in Fig. 1 but for E$^*$=150 MeV.
\item The same as in Fig. 1 but for E$^*$=250 MeV.
\item Emission widths for neutrons emitted from  $^{160}$Yb at E$^*$=150 MeV 
      and L=0, 20, 40 and 60$\hbar$ as a function of the elongation.
\item The same as in Fig. 4 but for protons.
\item The same as in Fig. 4 but for alpha particles.
\item Emission widths for neutrons emitted from different isotopes of Gd and Yb  
      at E$^*$=150 MeV and L=40$\hbar$ as a function of the elongation.
\item The same as in Fig. 7 but for protons.
\item The same as in Fig. 7 but for alpha particles.
\item Probabilities to emit a neutron with energy smaller than $e_n$ 
      from the fissioning nucleus $^{160}$Yb for three deformations 
      corresponding to the spherical shape, the top of the fission 
      barrier and the scission point.
\item The same as in Fig. 10 but for protons.
\item The same as in Fig. 10 but for alpha particles.
\item Probability to emit a neutron with energy smaller than 
      $e_n$ for three excitation energies (E$^*$=50, 150 and 250 MeV) 
      of the fissioning nucleus $^{160}$Yb.
\item The same as in Fig. 13 but for protons.
\item The same as in Fig. 13 but for alpha particles.
\item Number of trajectories out of sample of 10.000 leading to
      fission as a function of time with particle emission turned off.
      The calculation is done for three different temperatures 
      (T=3, 4 and 5 MeV) of the fissioning
      nucleus $^{160}$Yb. The heights of the corresponding fission barriers 
      (U$_B$) are indicated in the figure.
\item Multiplicity of the prefission neutrons (n), protons (p) and
      alpha particles ($\alpha$) as a function of time.
\item Ratio between the number of trajectories leading to
      fission and the total number of trajectories as a function of time
      for $^{160}$Yb. The emission of the light particles is taken into
      account.
\item Energy spectra of neutrons (n), protons (p) and alpha particles 
      emitted in coincidence with the fission events (solid lines). 
      The dashed lines correspond to the spectra of the particles emitted 
      from nuclei which end up as evaporation residua.
\item Yield of neutron emission $P_n$ as a function of the deformation 
      of the fissioning nucleus $^{160}$Yb with initial excitation energy 
      E$^*$=293 MeV and initial angular momentum L=52$\hbar$.
      The solid line corresponds to the trajectories which lead to
      fission while the dashed line to those leading to evaporation
      residua. The dotted line describes the distribution when the 
      emission of protons and alphas is turned off.
\item The same as in Fig. 20 but for L=45$\hbar$.
\item Multiplicity of  prefission neutrons (n), protons (p) and
      alpha particles ($\alpha$) as a function of angular momentum
      of the compound nucleus $^{160}$Yb with the initial
      excitation energy E$^*$=251 MeV.
\item The same as in Fig. 22 but for the excitation energy E$^*$=293 MeV.
\item Ratio between the number of trajectories leading to 
      fission and the total number of trajectories as a function of
      the initial angular momentum for two different excitation
      energies E$^*$=251 MeV and 293 MeV.
\item Fission barrier heights as a function of
      the initial angular momentum for two different excitation
      energies E$^*$=251 MeV and 293 MeV.
\item Multiplicities of prescission neutrons (n), protons (p)
      and alpha particles ($\alpha$) as a function of the strength
      of the friction force.
\end{enumerate}

\vfill\eject

\begin{center} {\large\bf Table 1} \end{center} 
 
\begin{table}[h] 
\begin{center} 

\begin{tabular}{|c|c|c|c|c|} 
\hline 
 & \multicolumn{2}{c|}{$E^*$=251 MeV}   &  \multicolumn{2}{c|}{$E^*$=293 MeV}\\ 
\hline 
 
 $\nu$   &   model &  exp.  &   model  &  exp. \\ 
\hline 
   n     &    5.98 & 6.10$\pm$ 1.5  &    7.80  & 8.50$\pm$ 1.6  \\ 
\hline 
   p     &    0.94 & 0.51$\pm$ 0.07   &    1.19  & 0.70$\pm$ 0.08  \\ 
\hline 
$\alpha$ &    0.58 & 0.48$\pm$ 0.07   &    0.66  & 0.75$\pm$ 0.08  \\ 
\hline 
\end{tabular}

\end{center} 
\end{table} 
 
\vfill\eject

\begin{figure}[t]\vspace*{-80mm} \hspace*{20mm}
\epsfxsize=150mm \epsfysize=200mm \epsfbox{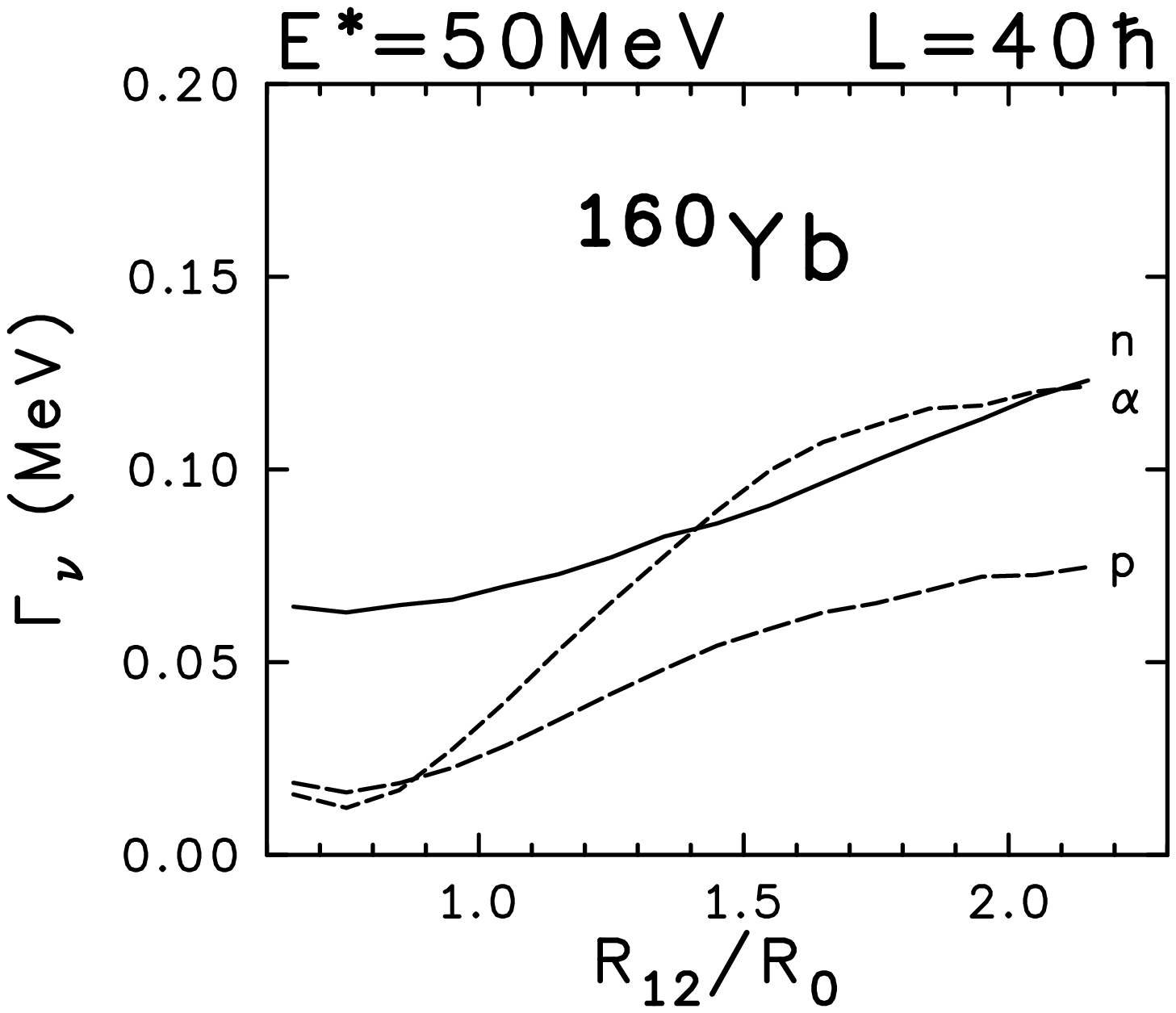}
\end{figure} 
\begin{center}\vspace*{4cm} Figure 1 \end{center}
\pagebreak[4]

\begin{figure}[t] \vspace*{-80mm} \hspace*{20mm}
\epsfxsize=150mm \epsfysize=200mm \epsfbox{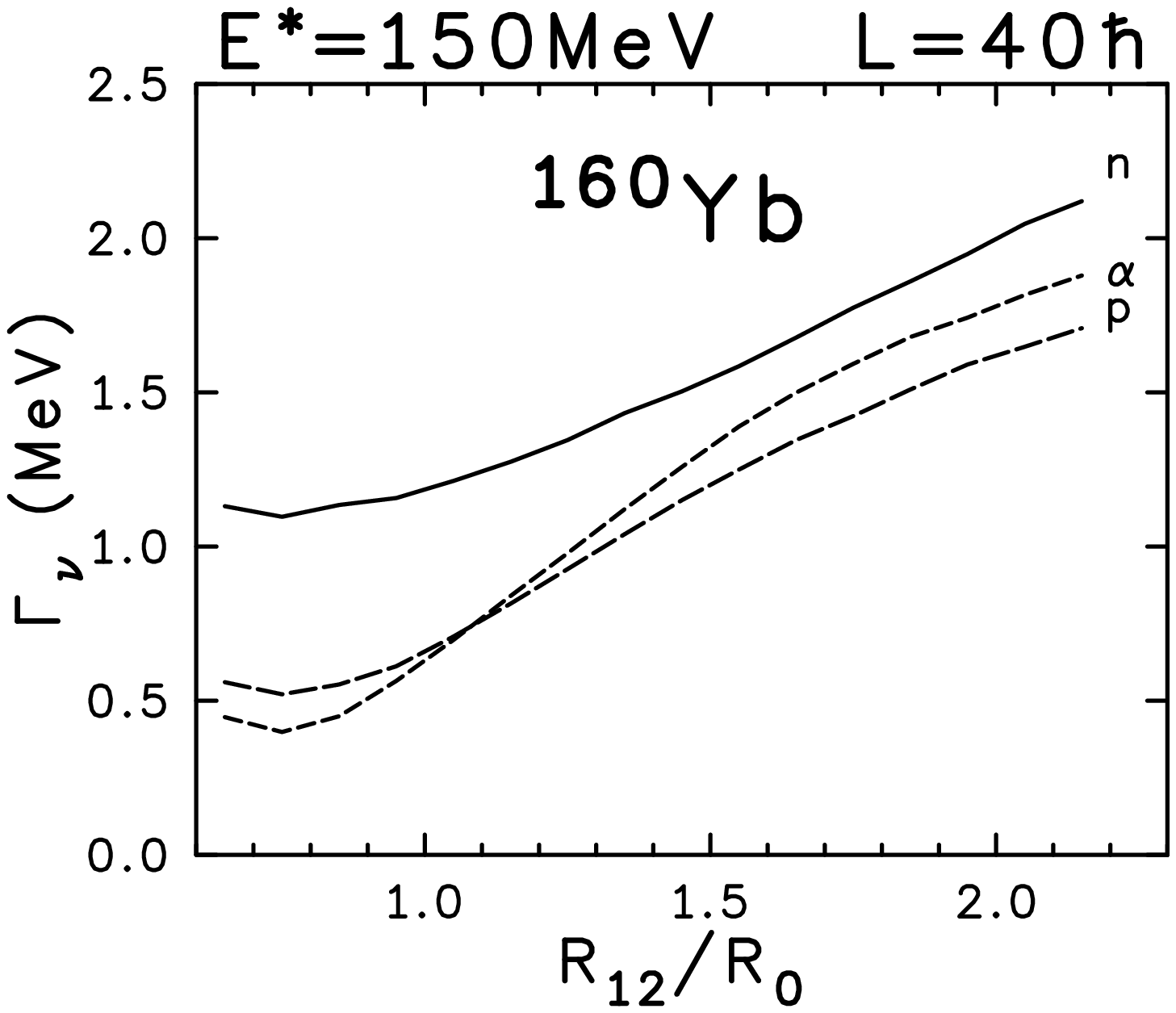}
\end{figure} 
\begin{center}\vspace*{4cm} Figure 2 \end{center}
\pagebreak[4]

\begin{figure}[t]\vspace*{-80mm} \hspace*{20mm}
\epsfxsize=150mm \epsfysize=200mm \epsfbox{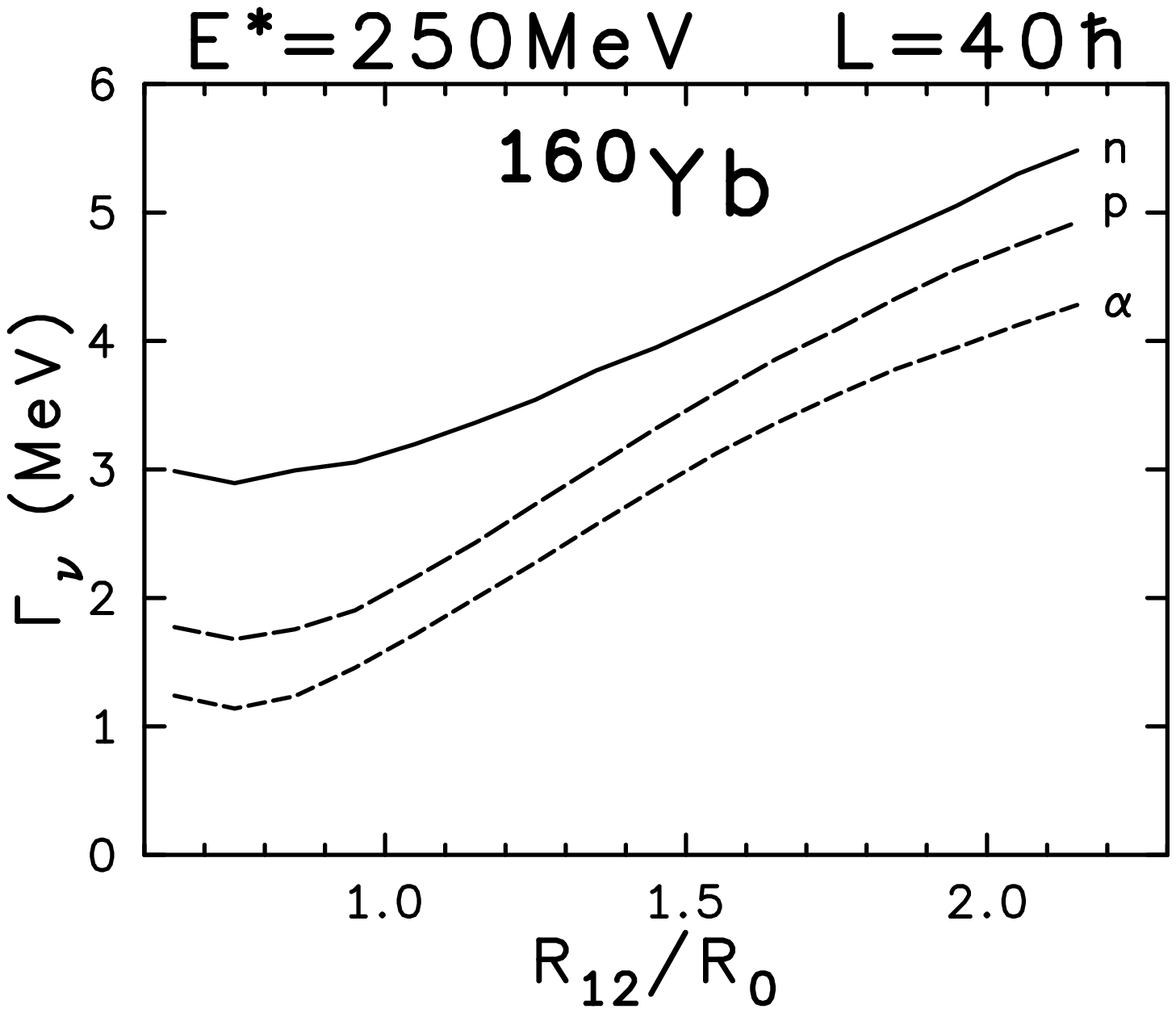}
\end{figure} 
\begin{center}\vspace*{4cm} Figure 3 \end{center}
\pagebreak[4]

\begin{figure}[t]\vspace*{-80mm} \hspace*{20mm}
\epsfxsize=150mm \epsfysize=200mm \epsfbox{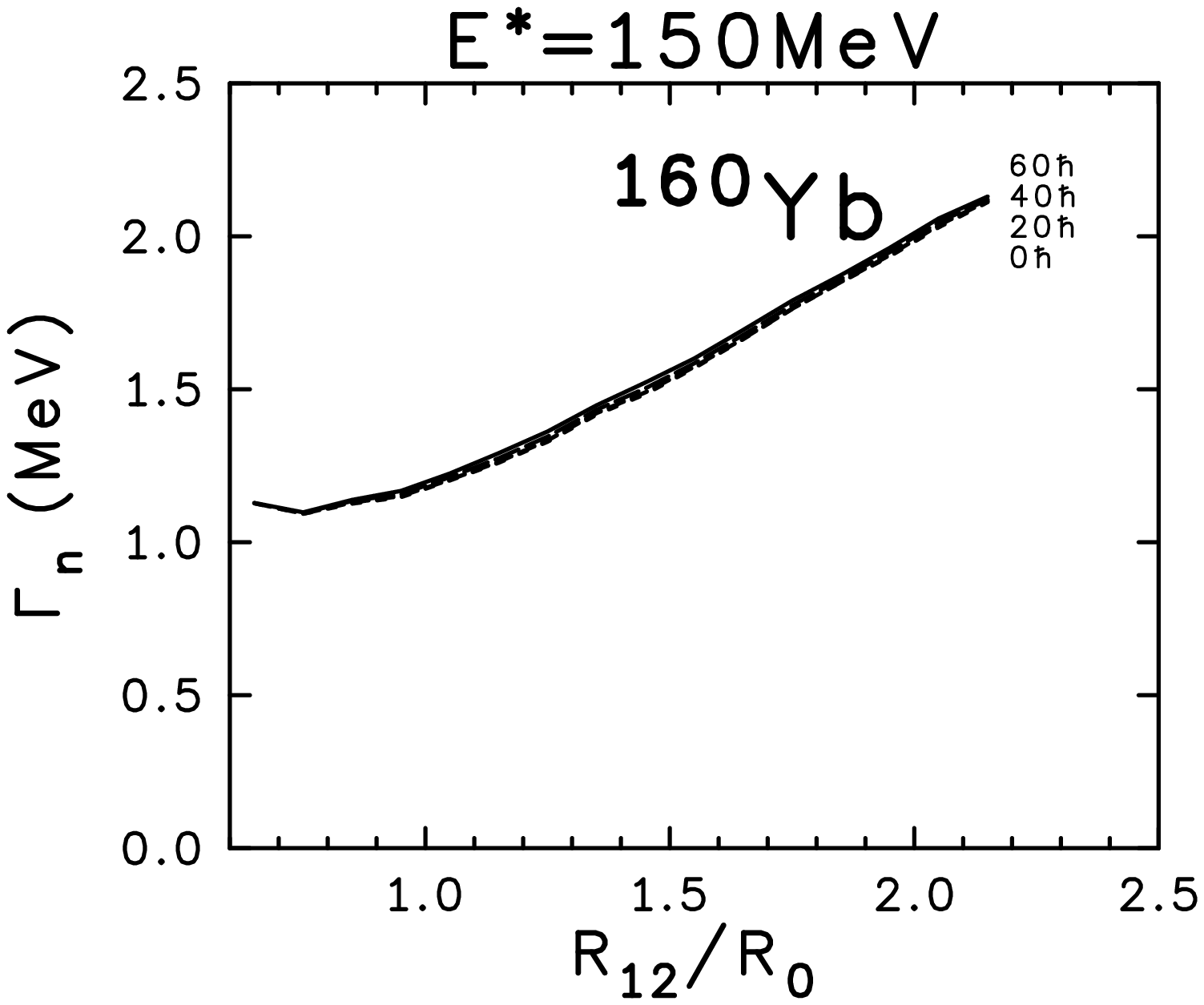}
\end{figure} 
\vfill
\begin{center}\vspace*{4cm} Figure 4 \end{center}
\pagebreak[4]

\begin{figure}[t]\vspace*{-80mm} \hspace*{20mm}
\epsfxsize=150mm \epsfysize=200mm \epsfbox{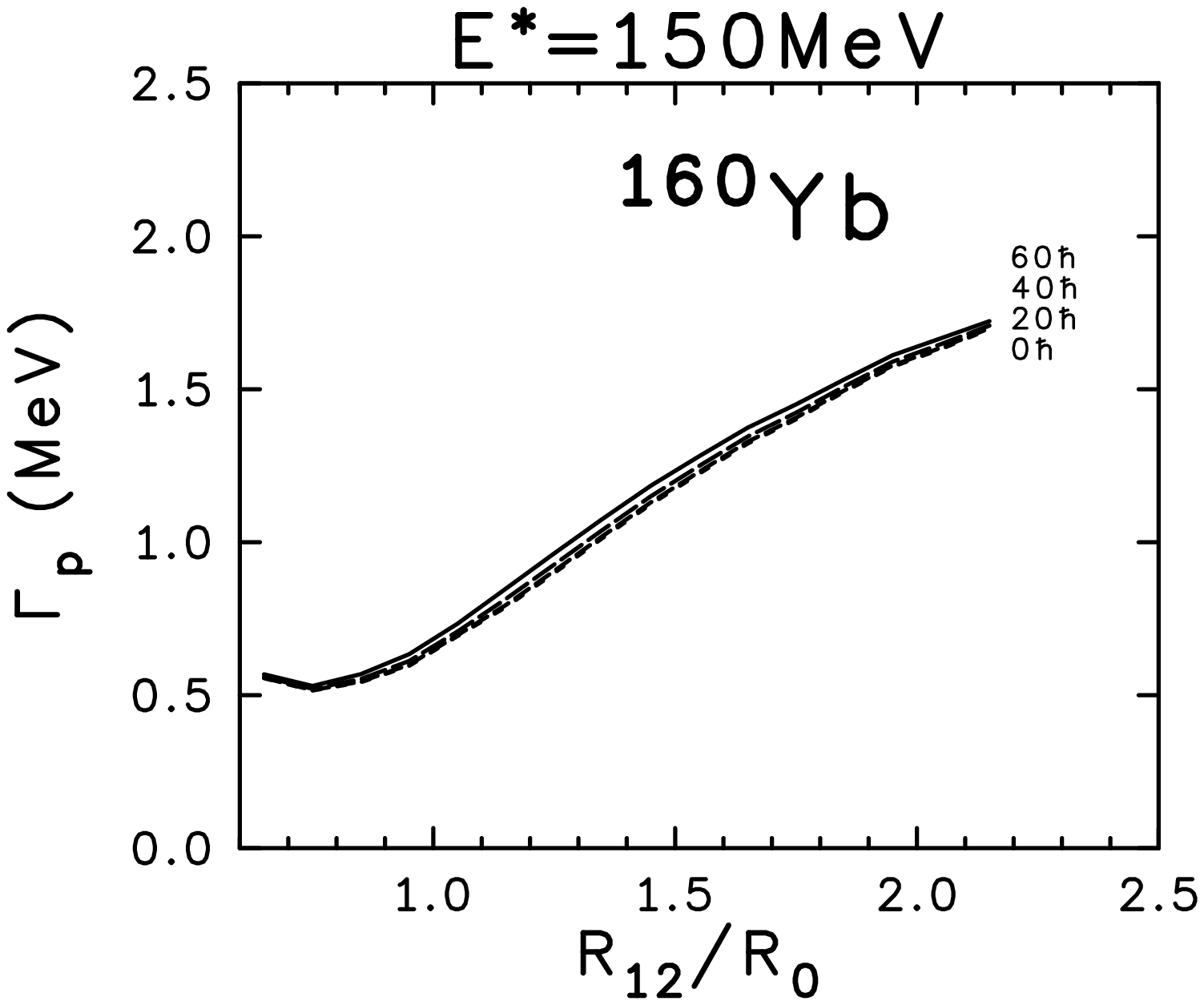}
\end{figure} 
\vfill
\begin{center}\vspace*{4cm} Figure 5 \end{center}
\pagebreak[4]

\begin{figure}[t]\vspace*{-80mm} \hspace*{20mm}
\epsfxsize=150mm \epsfysize=200mm \epsfbox{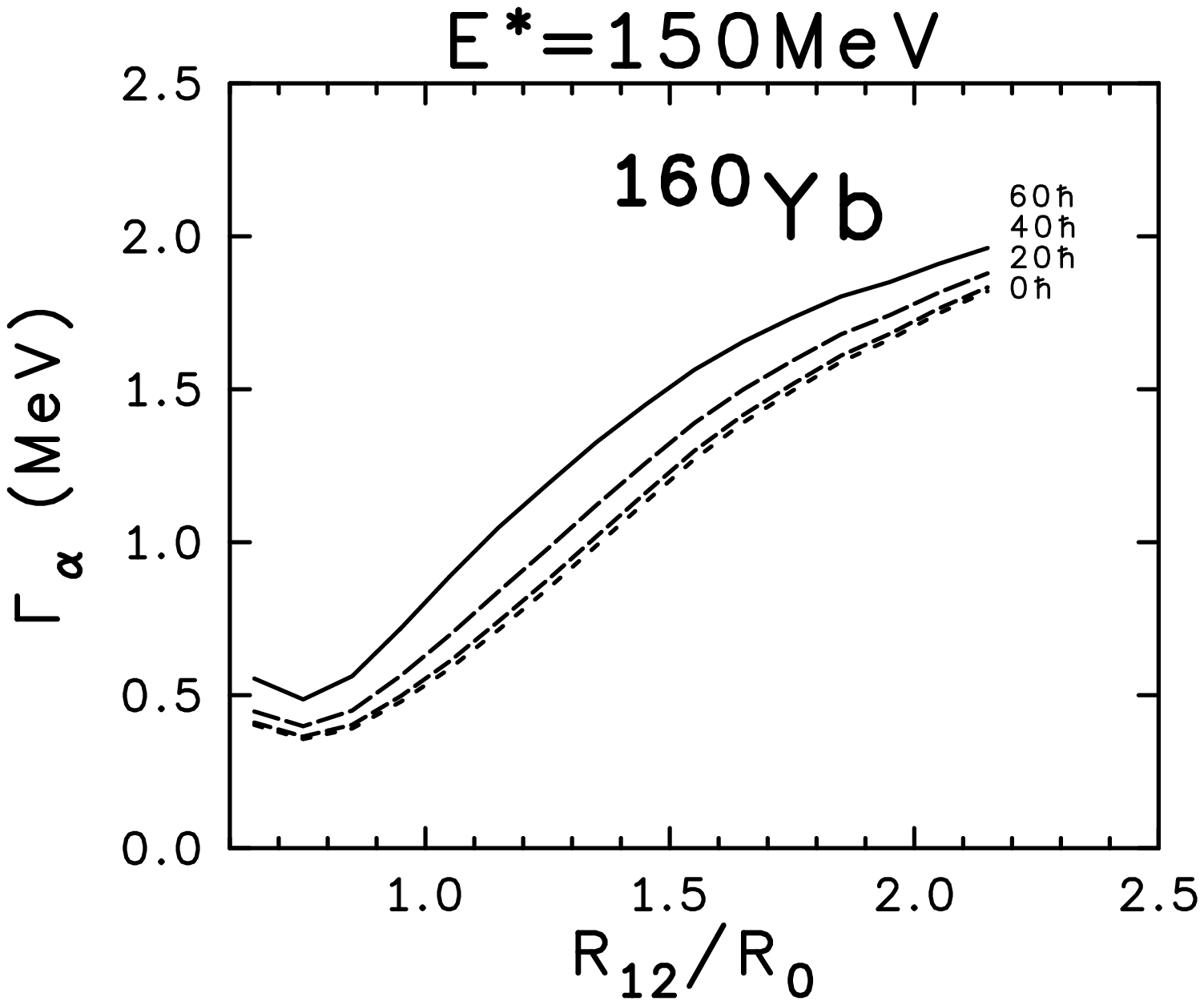}
\end{figure} 
\vfill
\begin{center}\vspace*{4cm} Figure 6 \end{center}
\pagebreak[4]

\begin{figure}[t]\vspace*{-80mm} \hspace*{20mm}
\epsfxsize=150mm \epsfysize=200mm \epsfbox{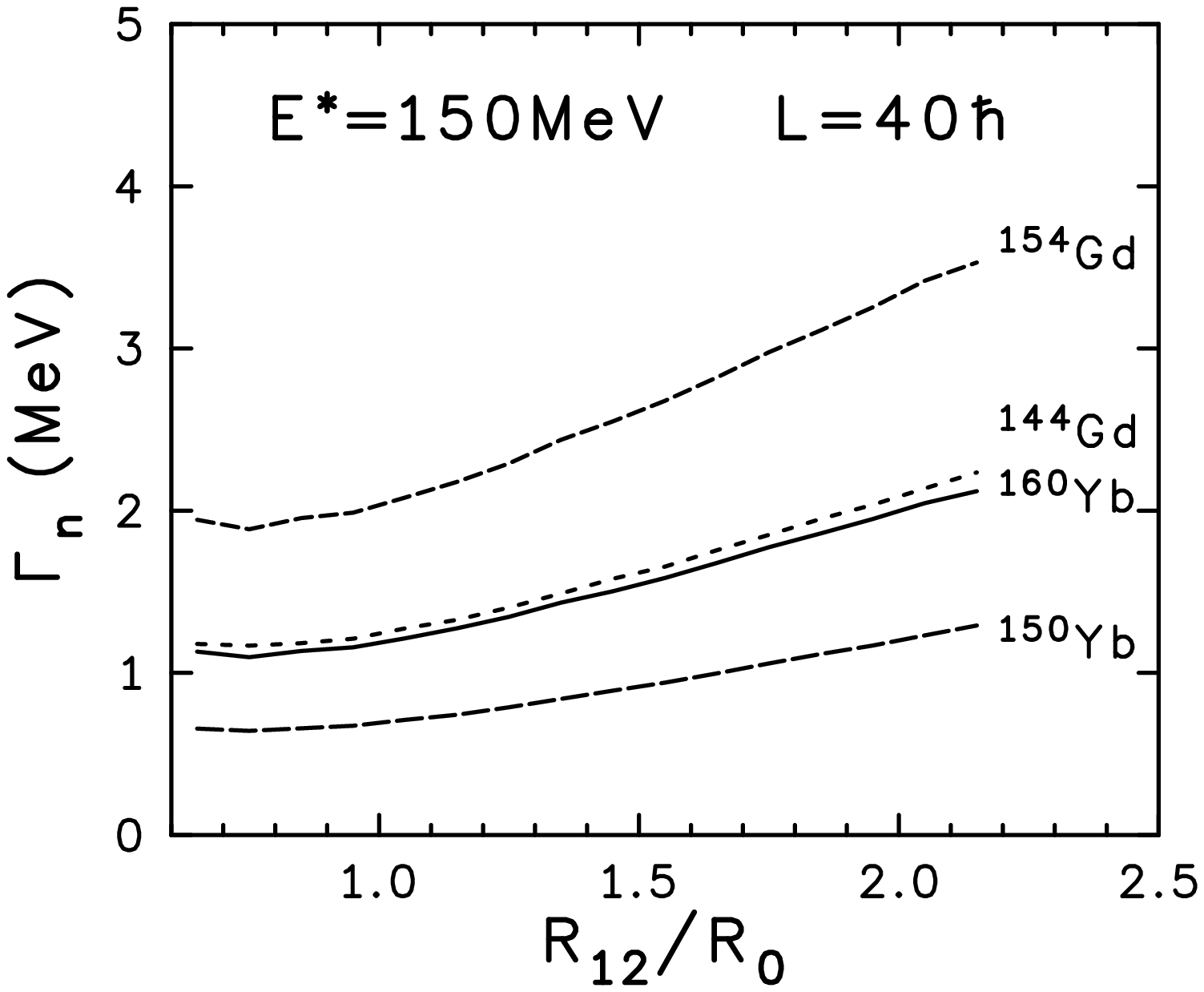}
\end{figure} 
\begin{center}\vspace*{4cm} Figure 7 \end{center}
\pagebreak[4]

\begin{figure}[t]\vspace*{-80mm} \hspace*{20mm}
\epsfxsize=150mm \epsfysize=200mm \epsfbox{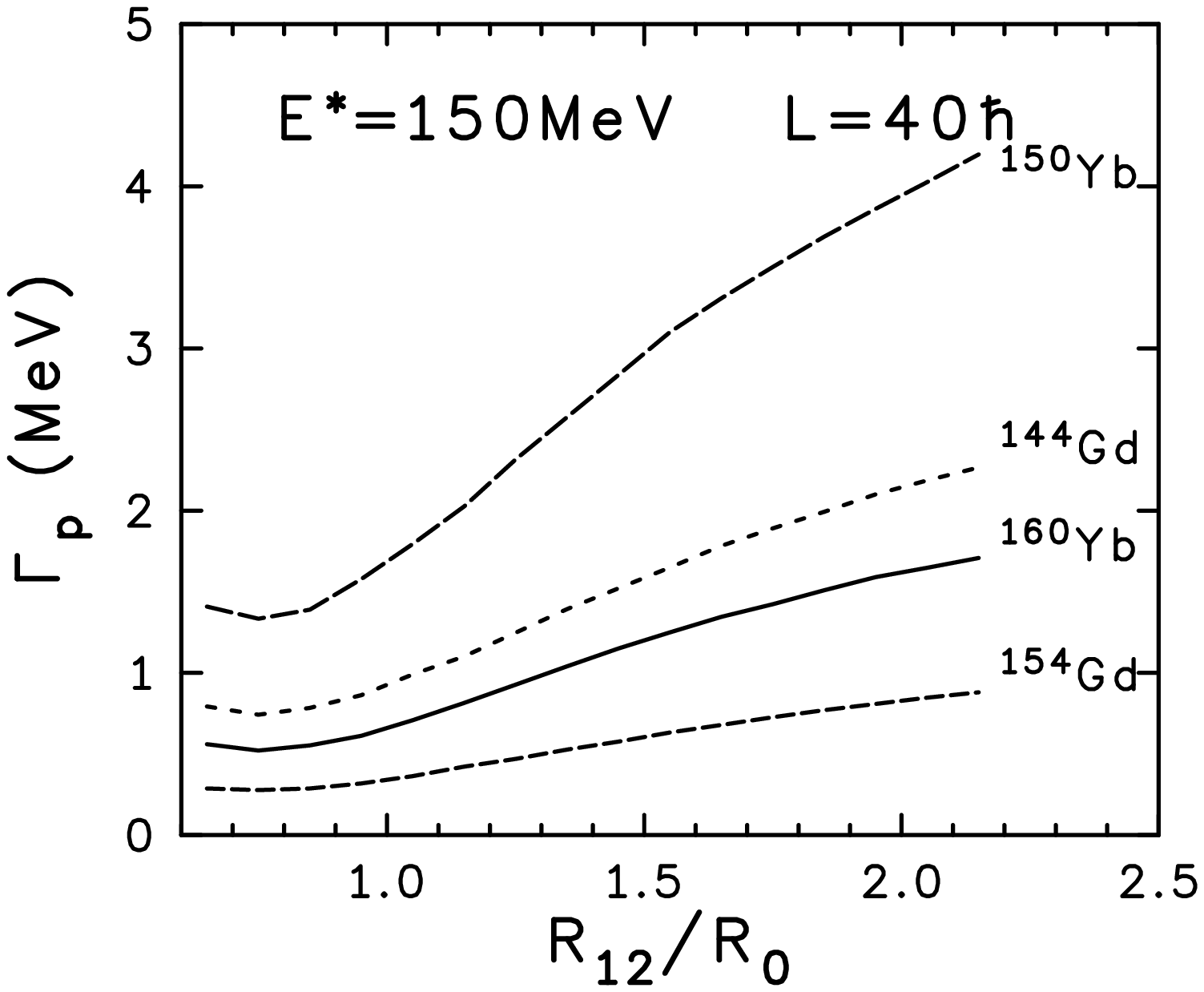}
\end{figure} 
\begin{center}\vspace*{4cm} Figure 8 \end{center}
\pagebreak[4]

\begin{figure}[t]\vspace*{-80mm} \hspace*{20mm}
\epsfxsize=150mm \epsfysize=200mm \epsfbox{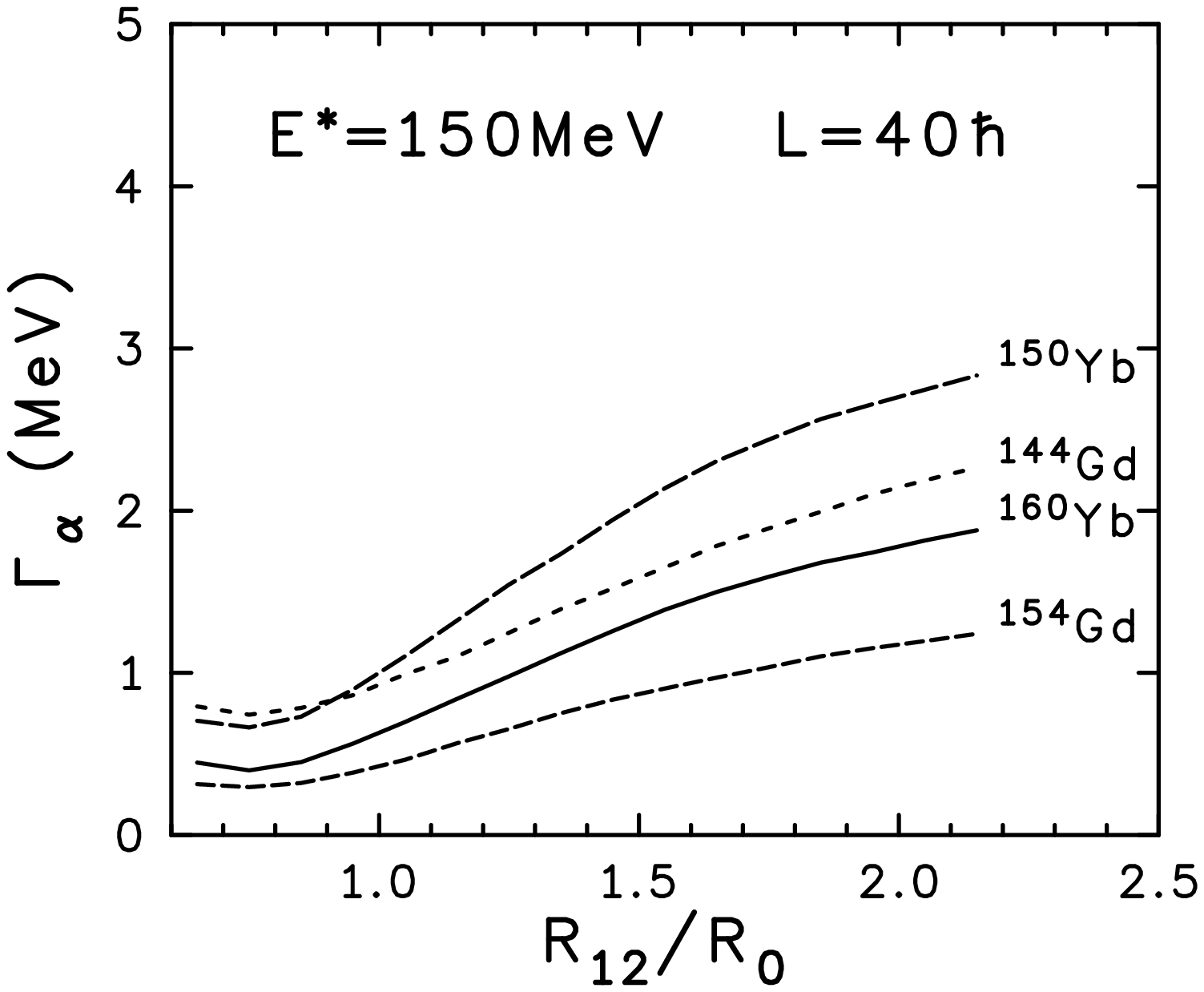}
\end{figure} 
\begin{center}\vspace*{4cm} Figure 9 \end{center}
\pagebreak[4]

\begin{figure}[t]\vspace*{-80mm} \hspace*{20mm}
\epsfxsize=150mm \epsfysize=200mm \epsfbox{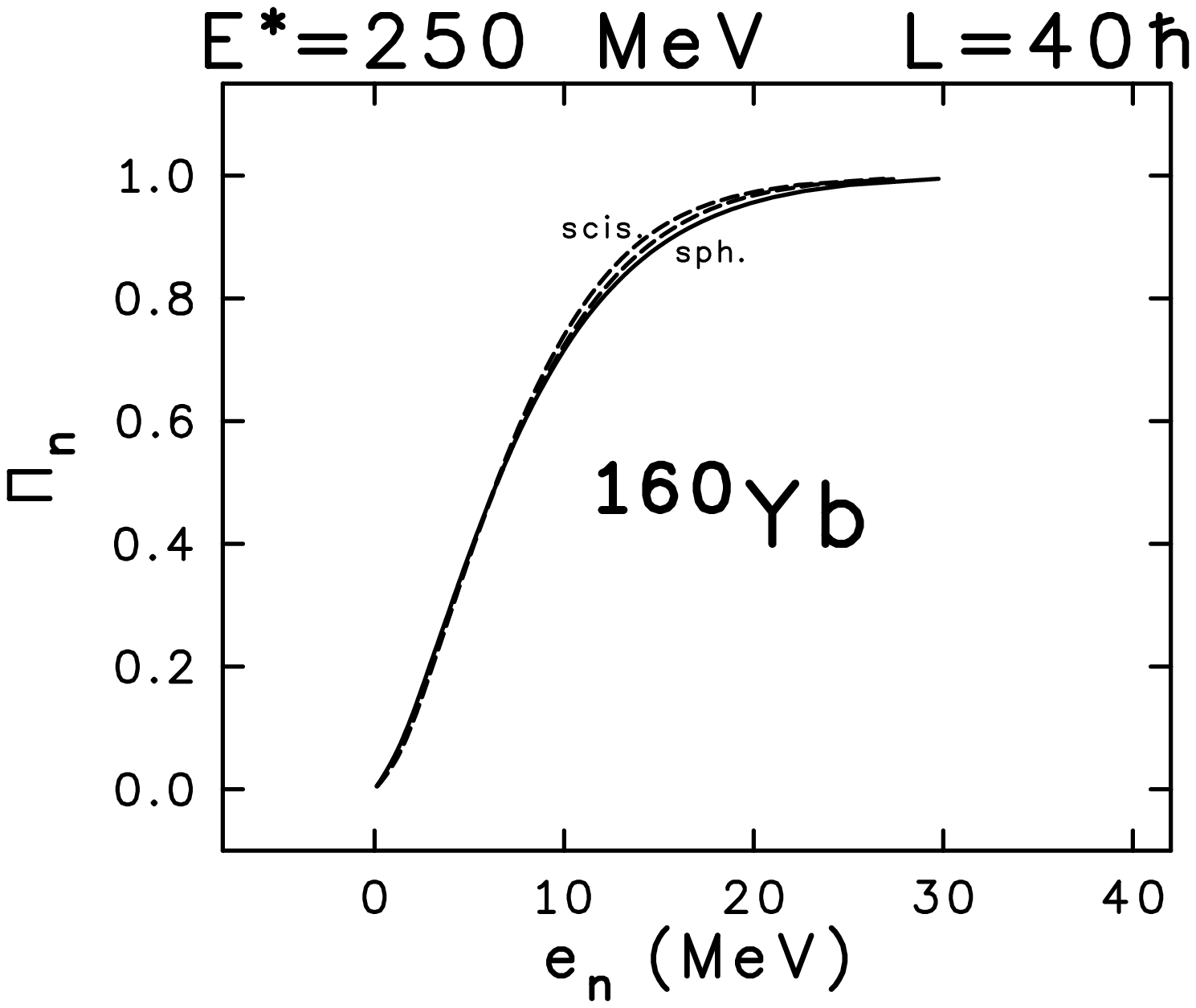}
\end{figure} 
\begin{center}\vspace*{4cm} Figure 10 \end{center}
\pagebreak[4]

\begin{figure}[t]\vspace*{-80mm} \hspace*{20mm}
\epsfxsize=150mm \epsfysize=200mm \epsfbox{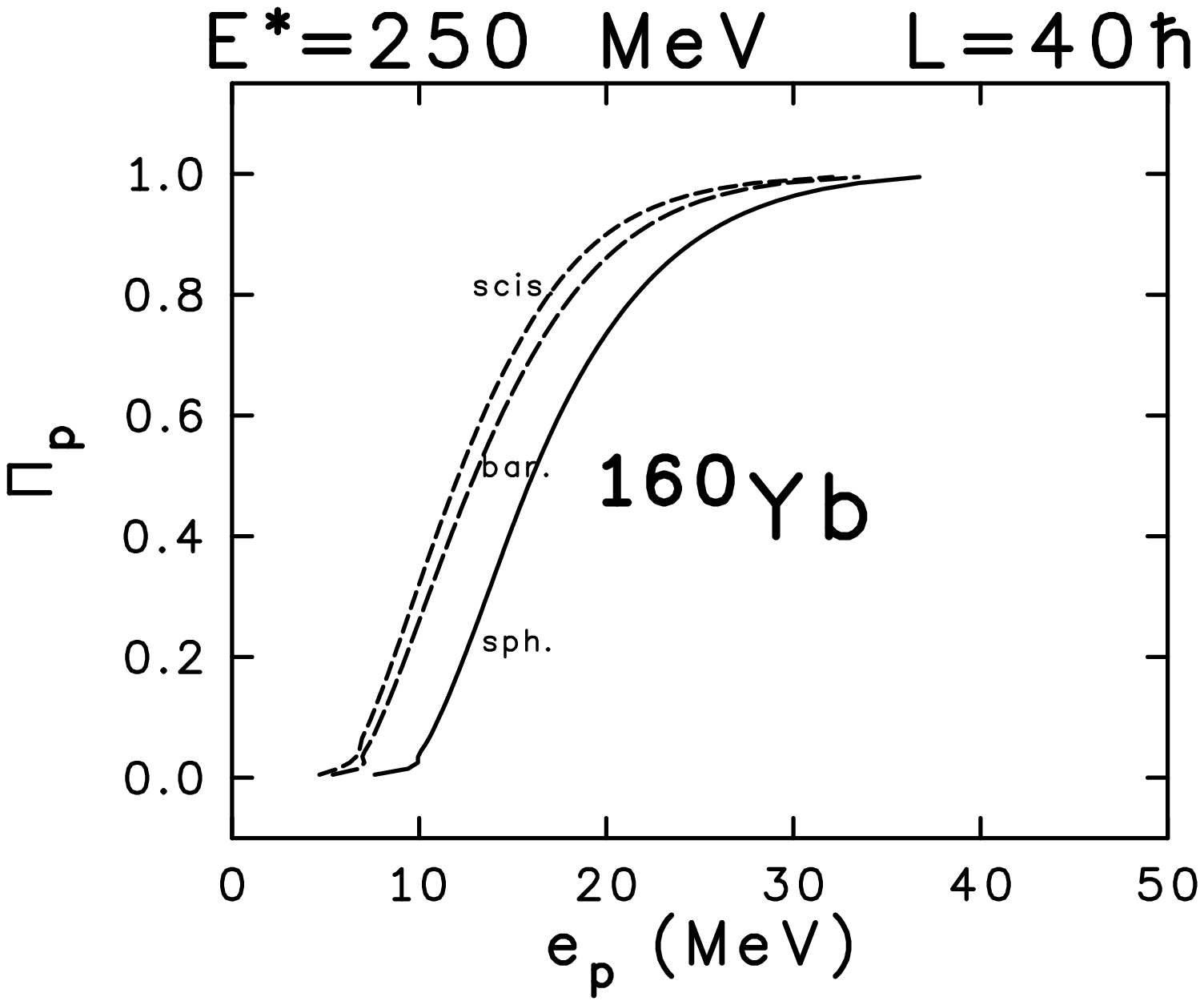}
\end{figure} 
\begin{center}\vspace*{4cm} Figure 11 \end{center}
\pagebreak[4]

\begin{figure}[t]\vspace*{-80mm} \hspace*{20mm}
\epsfxsize=150mm \epsfysize=200mm \epsfbox{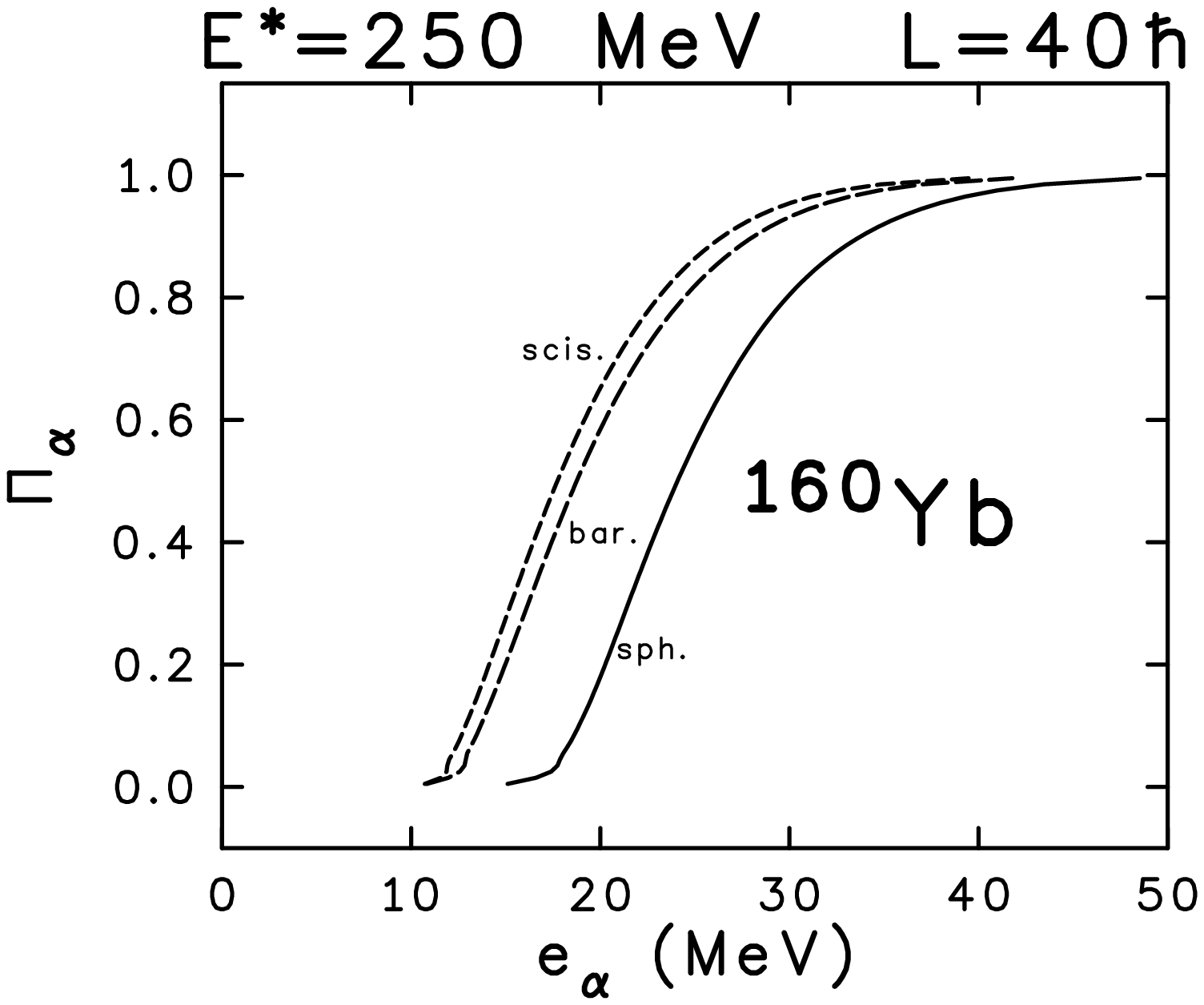}
\end{figure} 
\begin{center}\vspace*{4cm} Figure 12 \end{center}
\pagebreak[4]

\begin{figure}[t]\vspace*{-80mm} \hspace*{20mm}
\epsfxsize=150mm \epsfysize=200mm \epsfbox{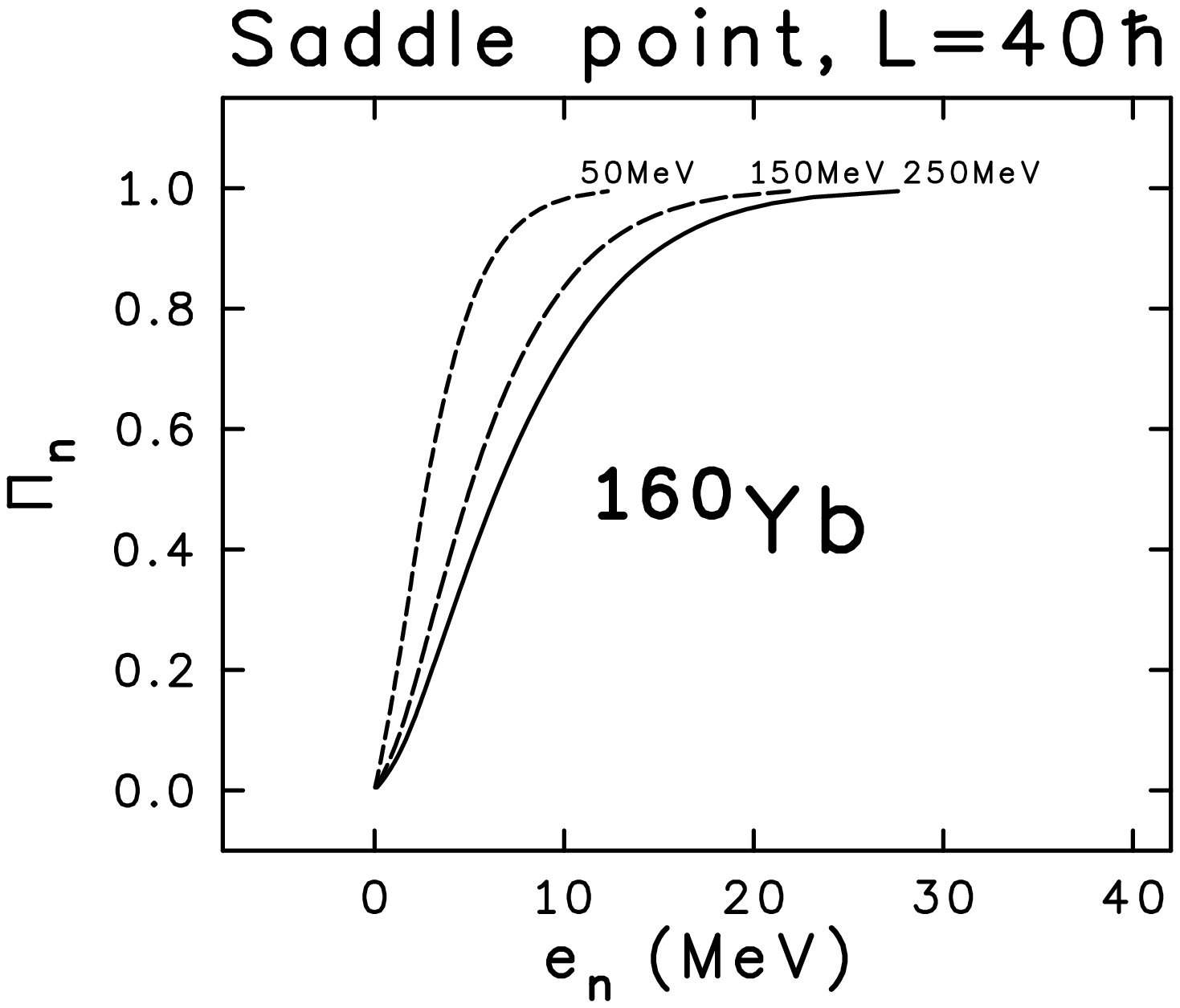}
\end{figure} 
\begin{center}\vspace*{4cm} Figure 13 \end{center}
\pagebreak[4]

\begin{figure}[t]\vspace*{-80mm} \hspace*{20mm}
\epsfxsize=150mm \epsfysize=200mm \epsfbox{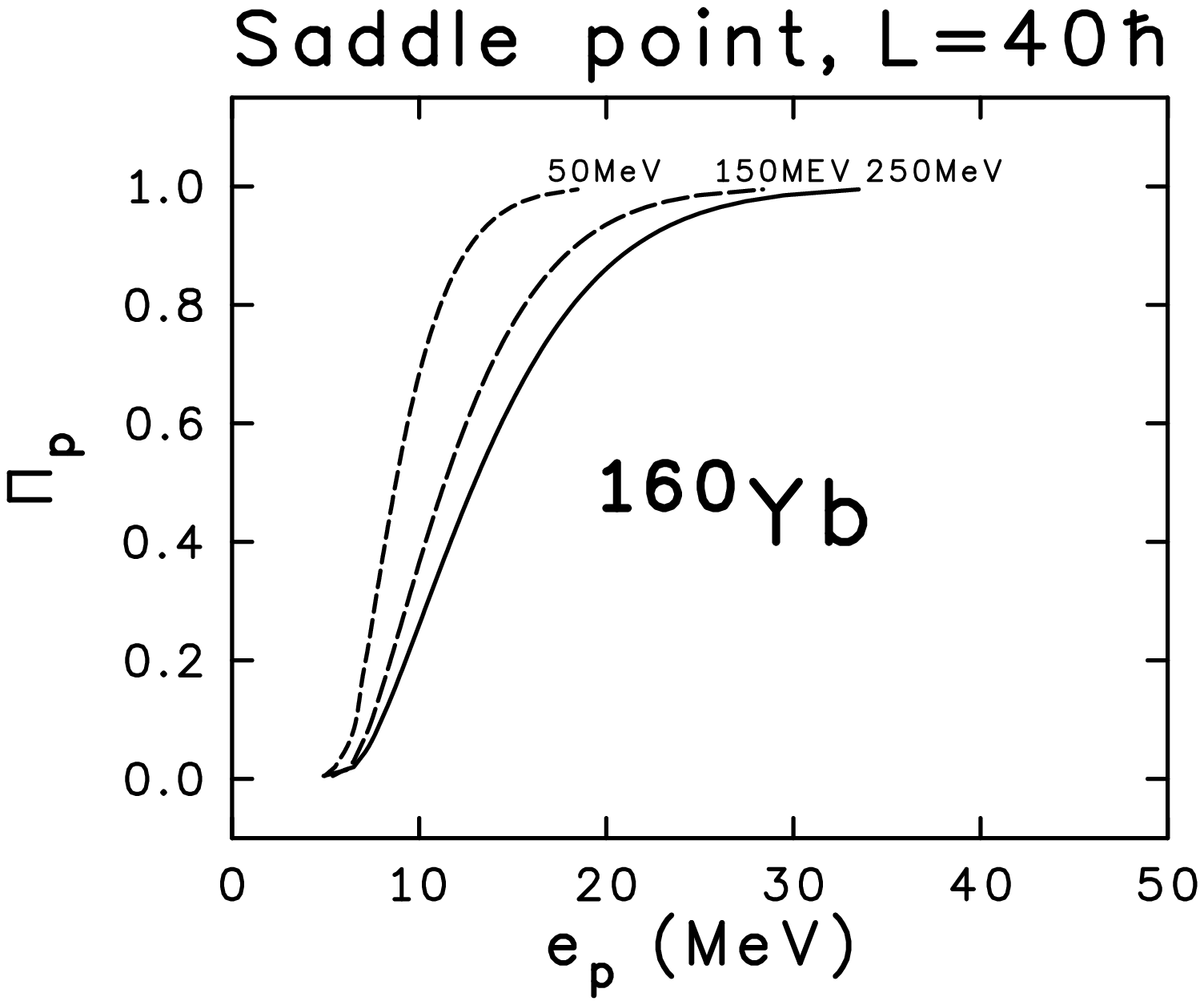}
\end{figure} 
\begin{center}\vspace*{4cm} Figure 14 \end{center}
\pagebreak[4]

\begin{figure}[t]\vspace*{-80mm} \hspace*{20mm}
\epsfxsize=150mm \epsfysize=200mm \epsfbox{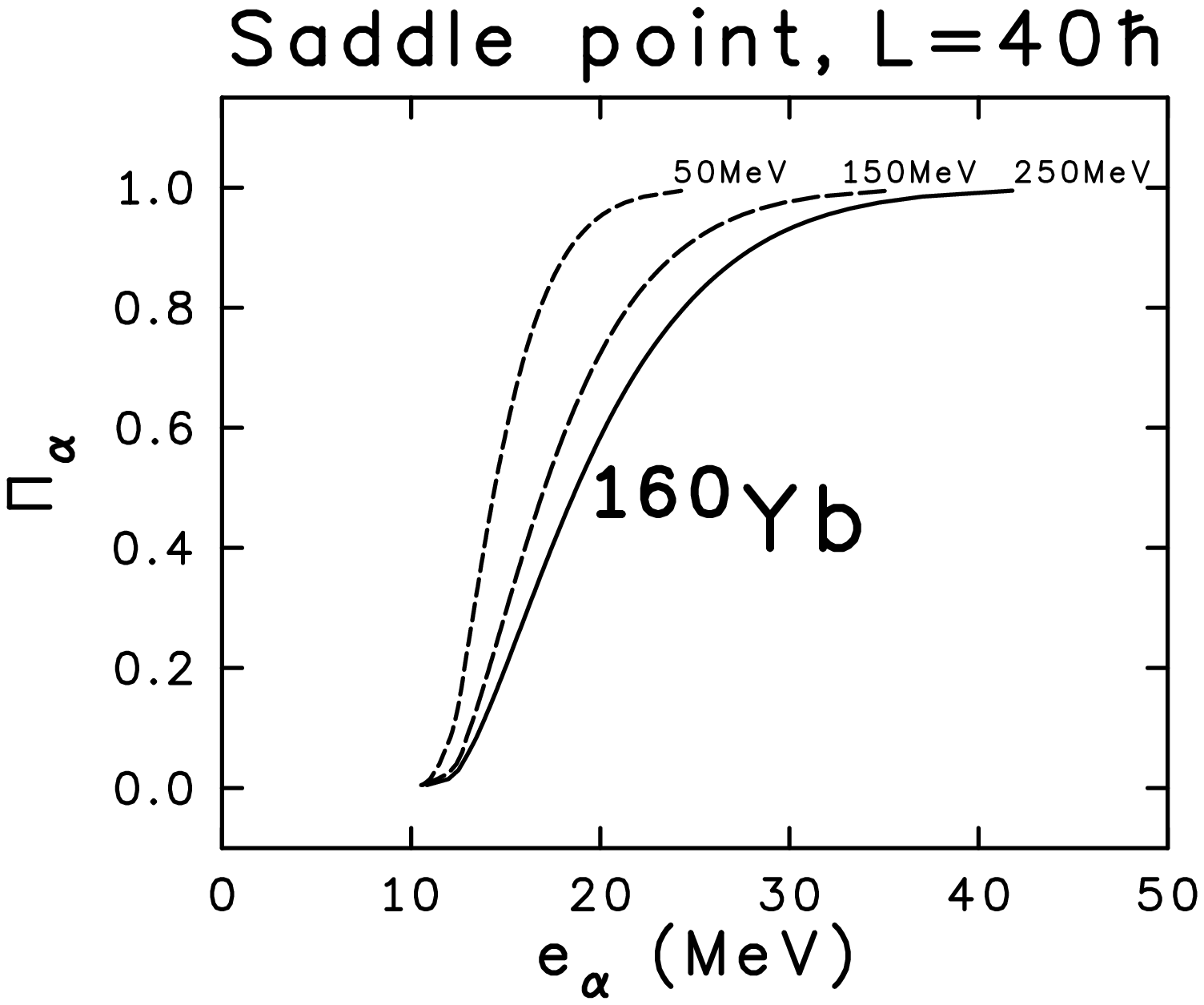}
\end{figure} 
\begin{center}\vspace*{4cm} Figure 15 \end{center}
\pagebreak[4]

\begin{figure}[t]\vspace*{-80mm} \hspace*{20mm}
\epsfxsize=150mm \epsfysize=200mm \epsfbox{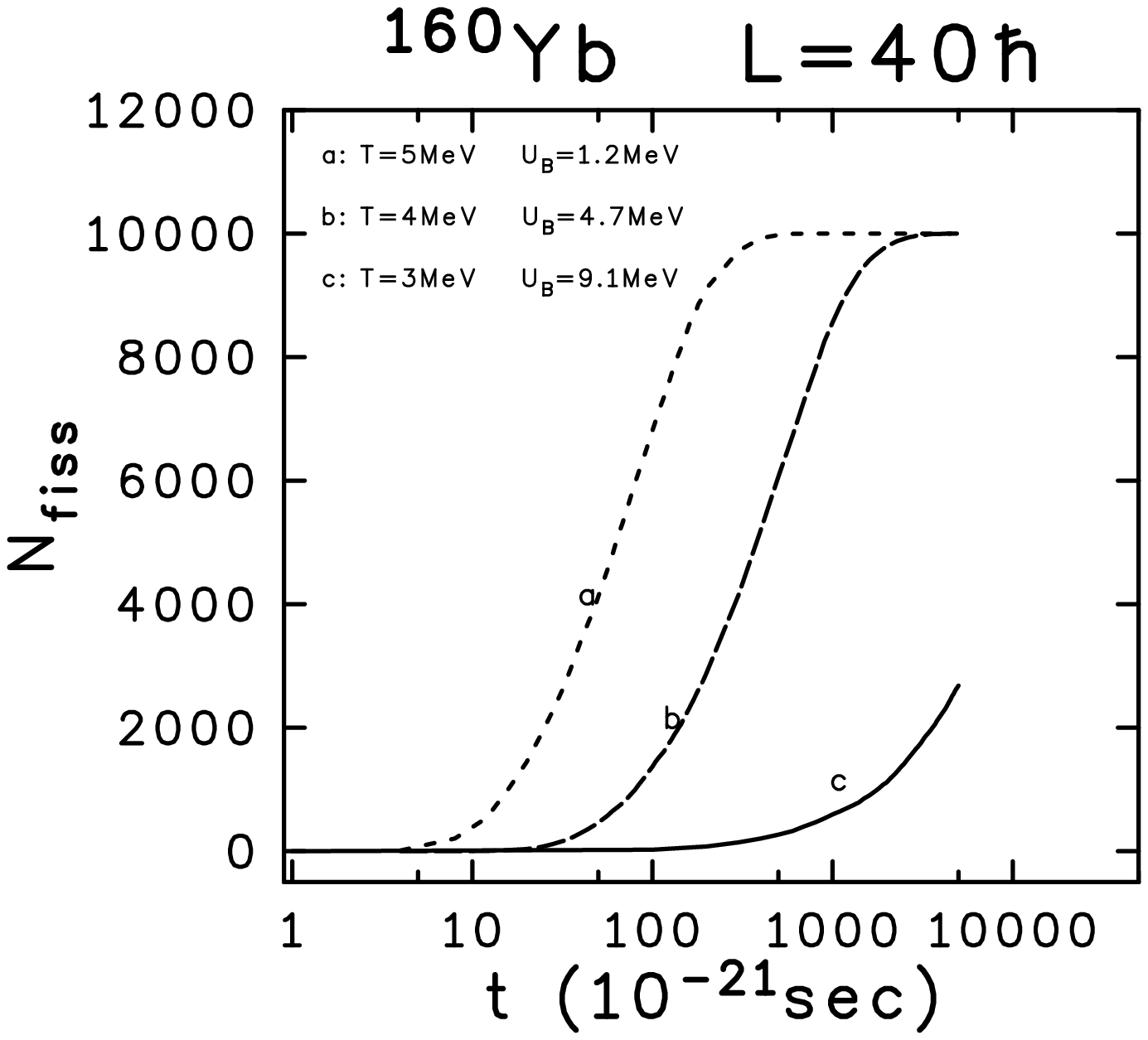}
\end{figure} 
\begin{center}\vspace*{4cm} Figure 16 \end{center}
\pagebreak[4]

\begin{figure}[t]\vspace*{-80mm} \hspace*{20mm}
\epsfxsize=150mm \epsfysize=200mm \epsfbox{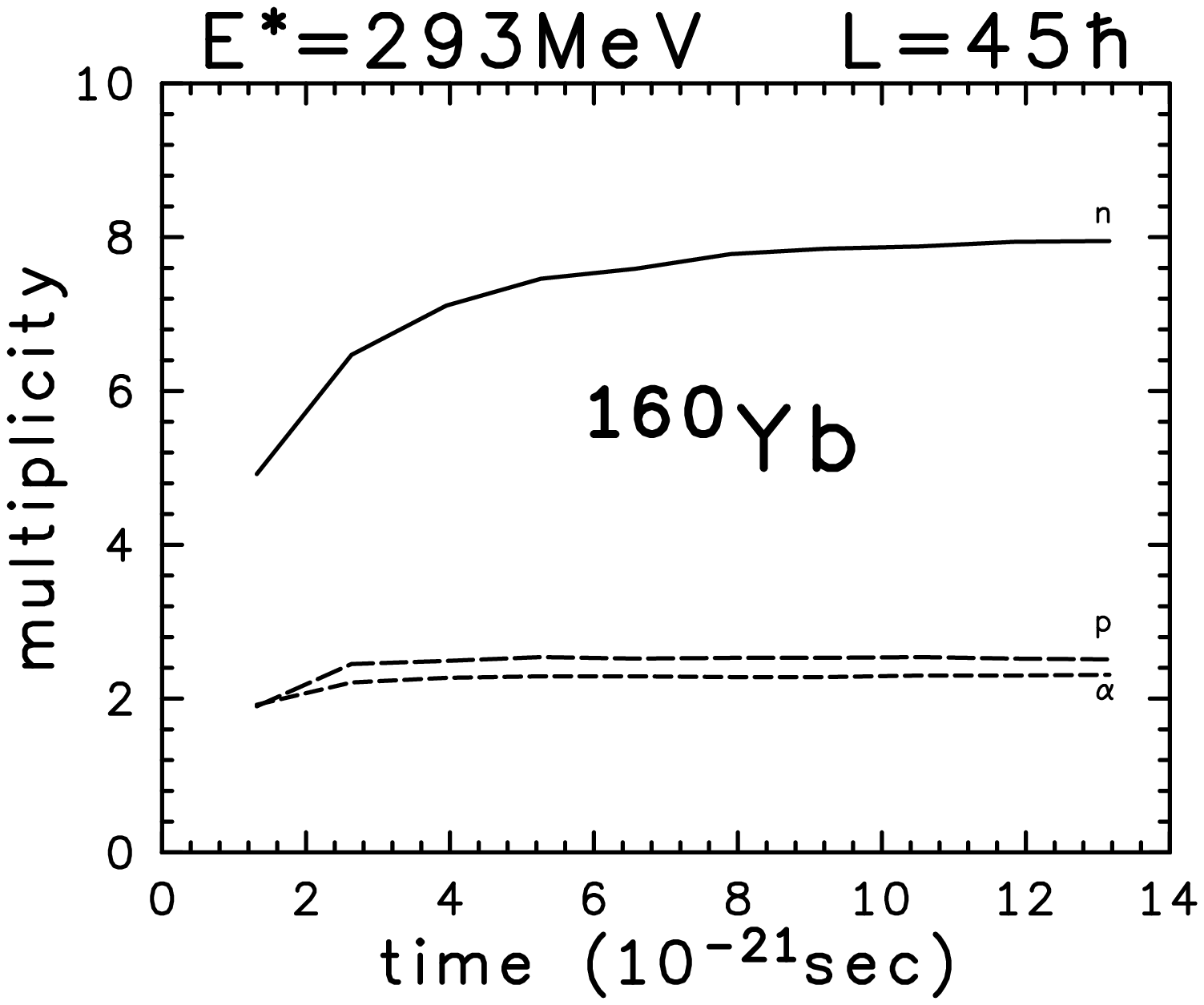}
\end{figure} 
\begin{center}\vspace*{4cm} Figure 17 \end{center}
\pagebreak[4]

\begin{figure}[t]\vspace*{-80mm} \hspace*{20mm}
\epsfxsize=150mm \epsfysize=200mm \epsfbox{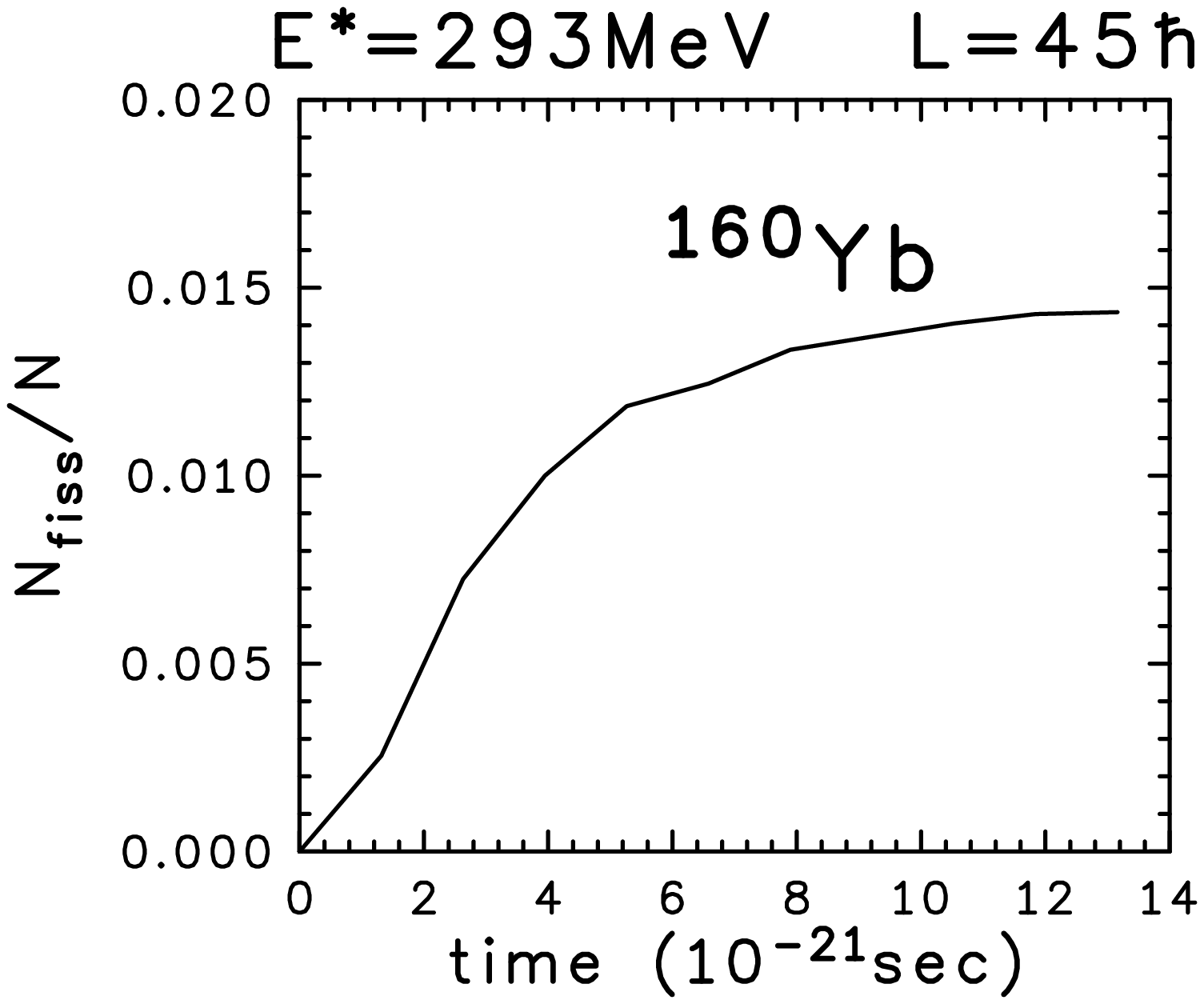}
\end{figure} 
\begin{center}\vspace*{4cm} Figure 18 \end{center} 
\pagebreak[4]

\begin{figure}[t]\vspace*{-80mm} \hspace*{20mm}
\epsfxsize=150mm \epsfysize=200mm \epsfbox{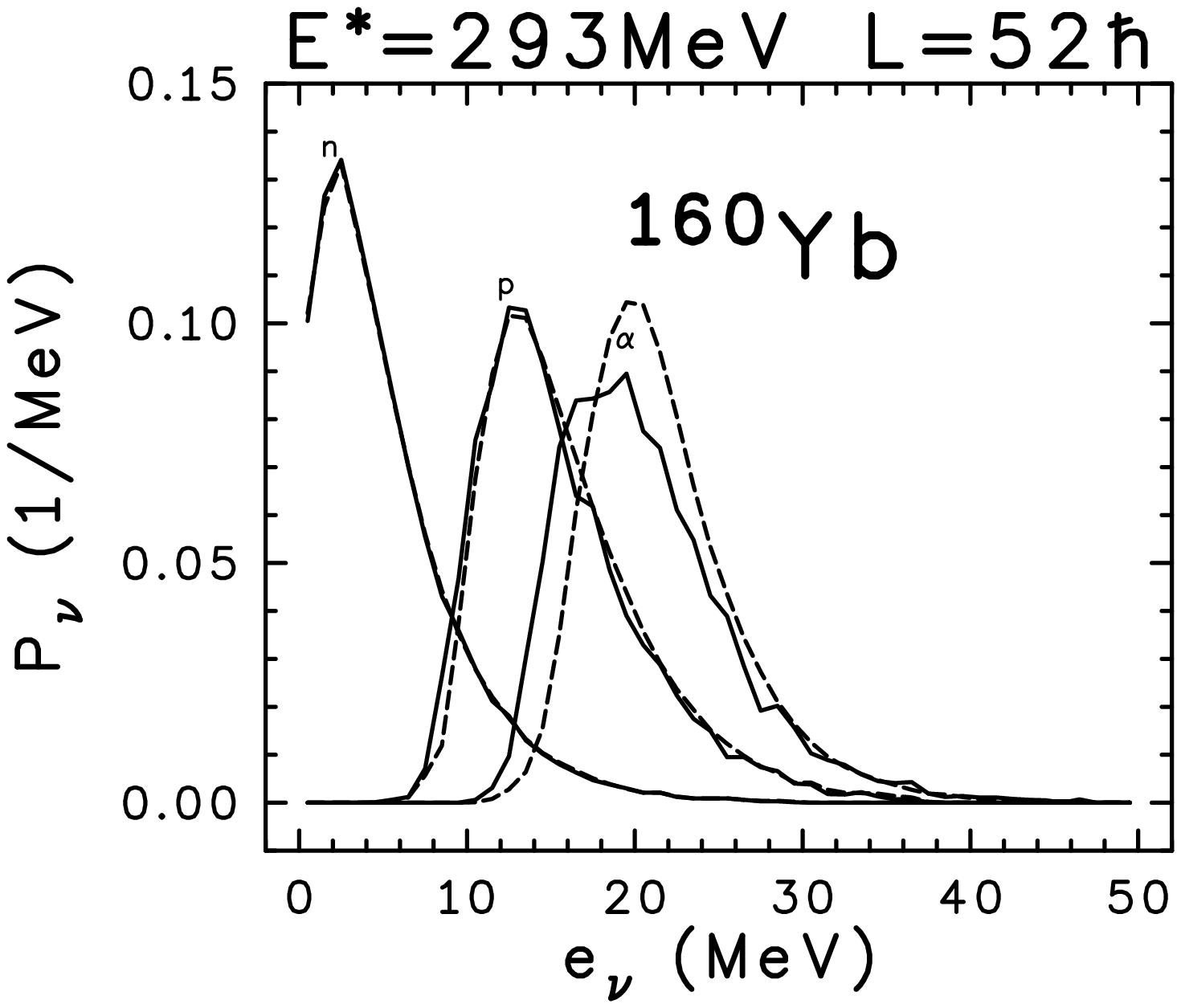}
\end{figure} 
\begin{center}\vspace*{4cm} Figure 19 \end{center}
\pagebreak[4]

\begin{figure}[t]\vspace*{-80mm} \hspace*{20mm}
\epsfxsize=150mm \epsfysize=200mm \epsfbox{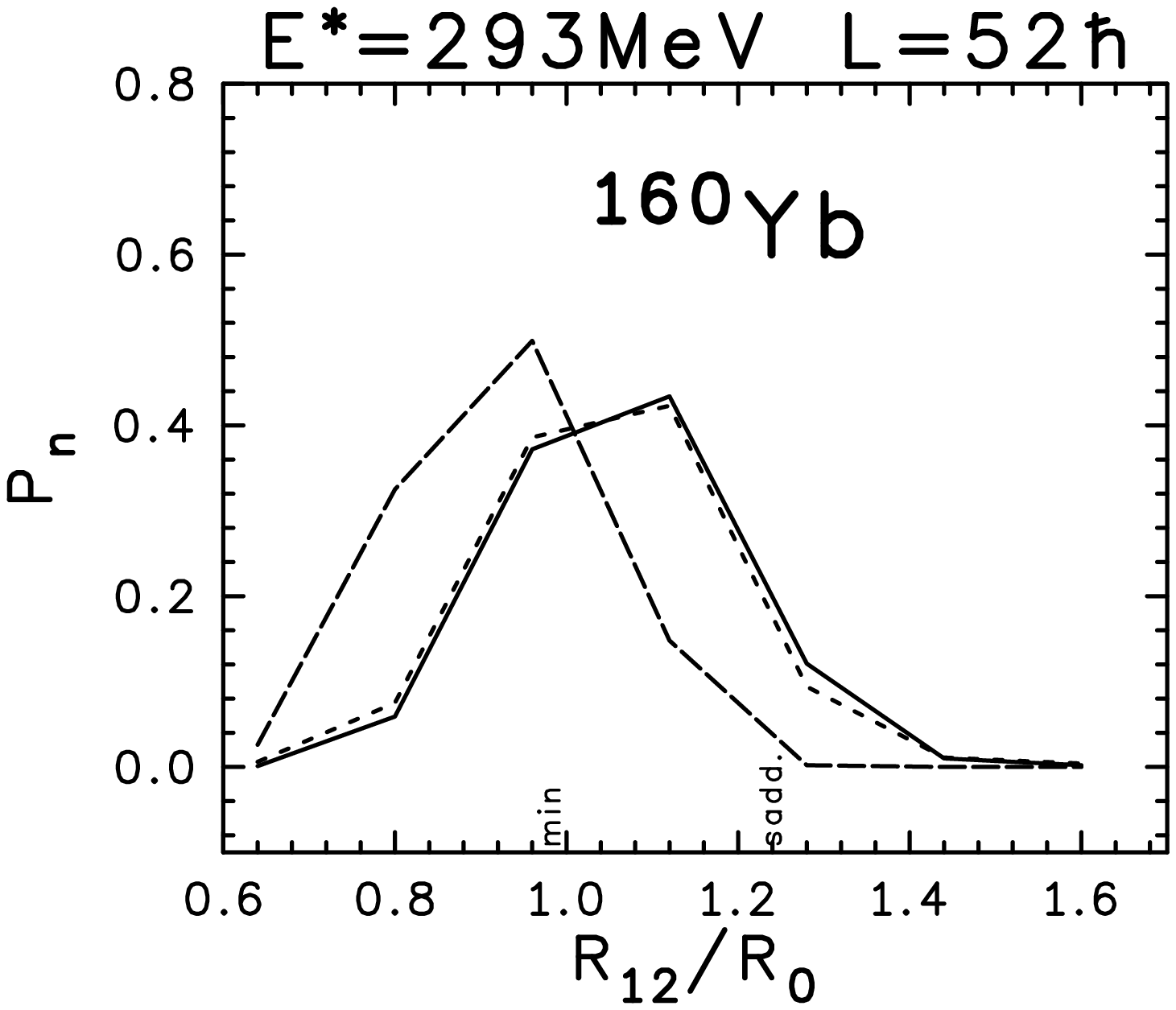}
\end{figure} 
\begin{center}\vspace*{4cm} Figure 20 \end{center}
\pagebreak[4]

\begin{figure}[t]\vspace*{-80mm} \hspace*{20mm}
\epsfxsize=150mm \epsfysize=200mm \epsfbox{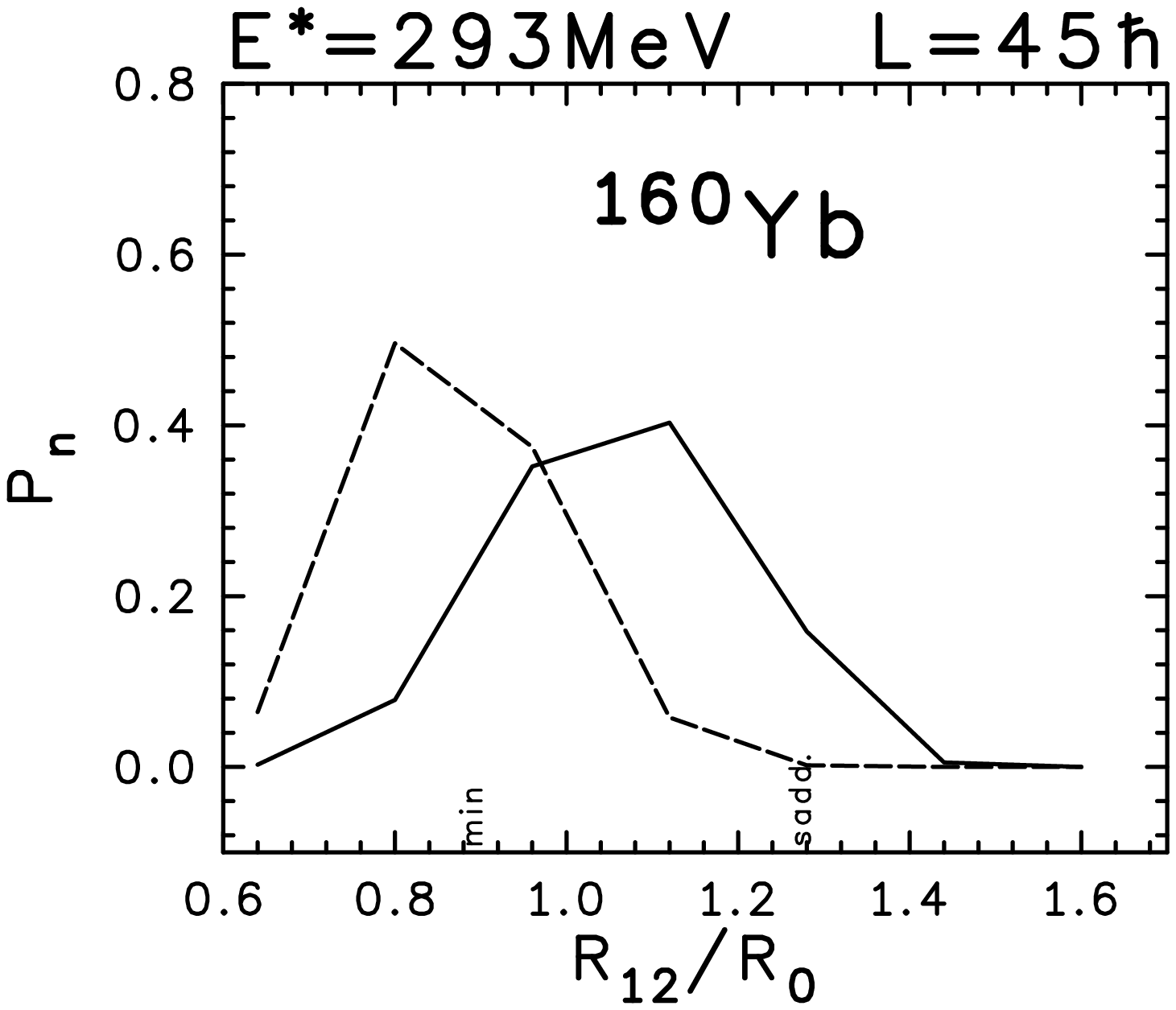}
\end{figure} 
\begin{center}\vspace*{4cm} Figure 21 \end{center}
\pagebreak[4]

\begin{figure}[t]\vspace*{-80mm} \hspace*{20mm}
\epsfxsize=150mm \epsfysize=200mm \epsfbox{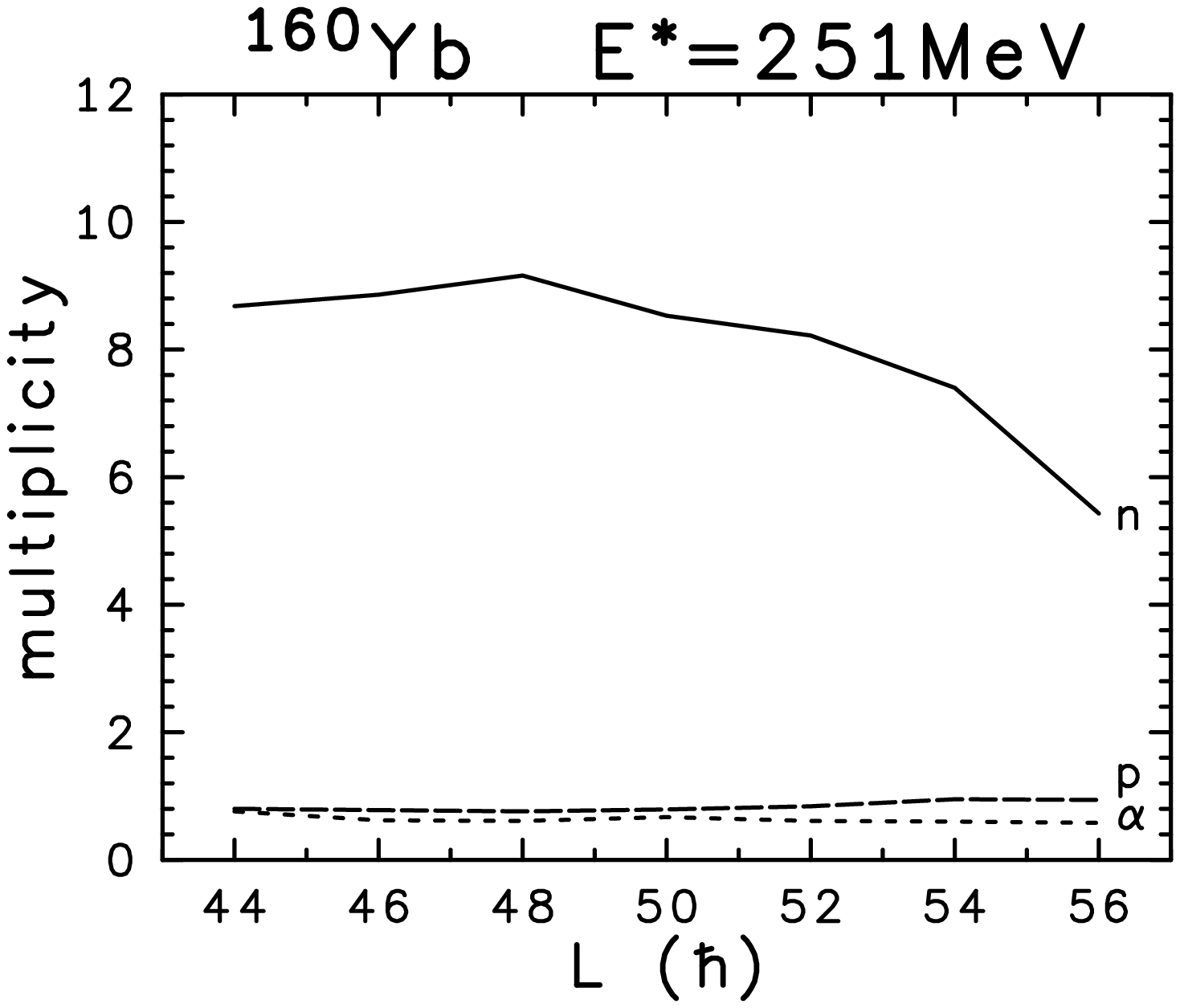}
\end{figure} 
\begin{center}\vspace*{4cm} Figure 22 \end{center}
\pagebreak[4]

\begin{figure}[t]\vspace*{-80mm} \hspace*{20mm}
\epsfxsize=150mm \epsfysize=200mm \epsfbox{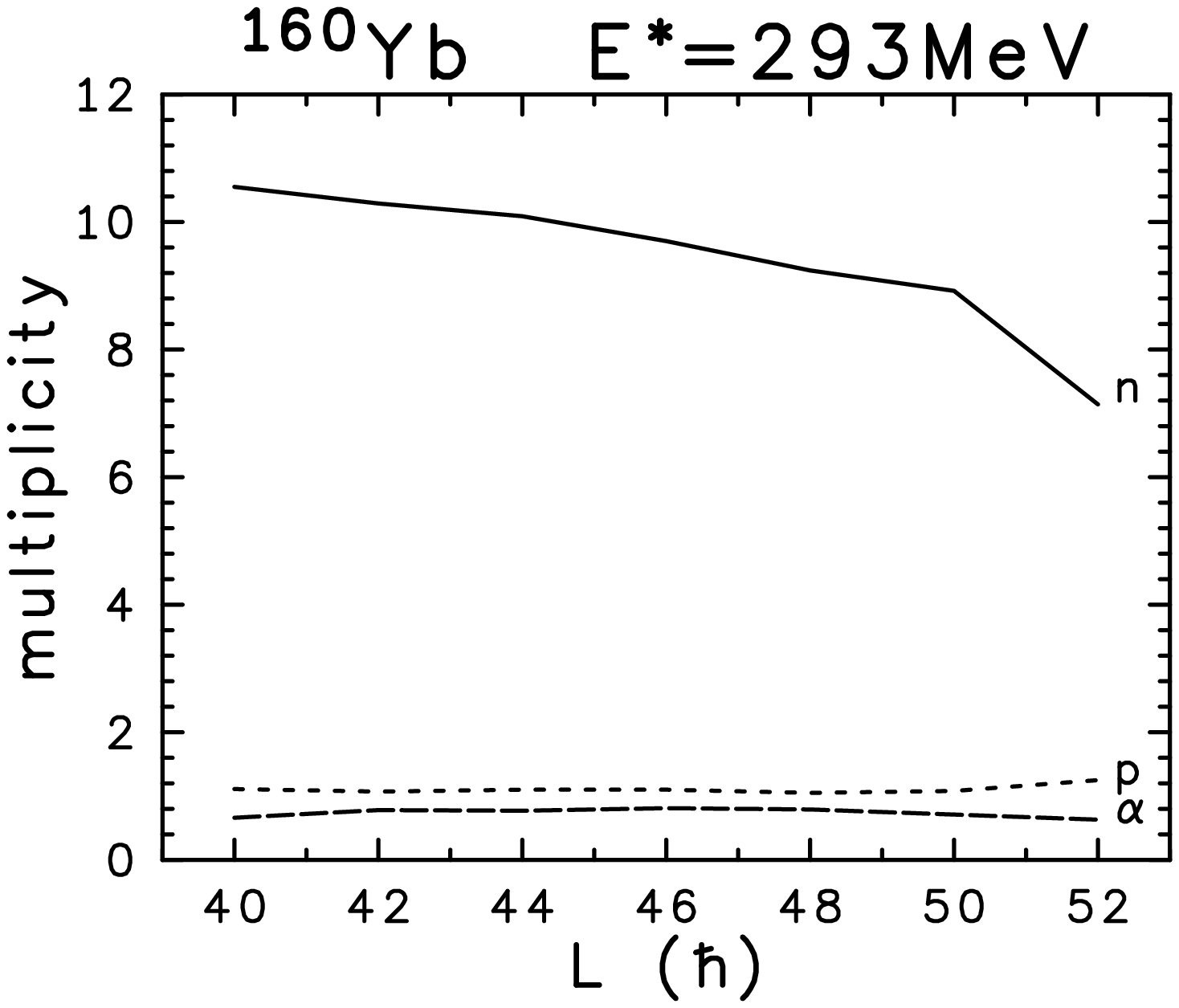}
\end{figure} 
\begin{center}\vspace*{4cm} Figure 23 \end{center}
\pagebreak[4]

\begin{figure}[t]\vspace*{-80mm} \hspace*{20mm}
\epsfxsize=150mm \epsfysize=200mm \epsfbox{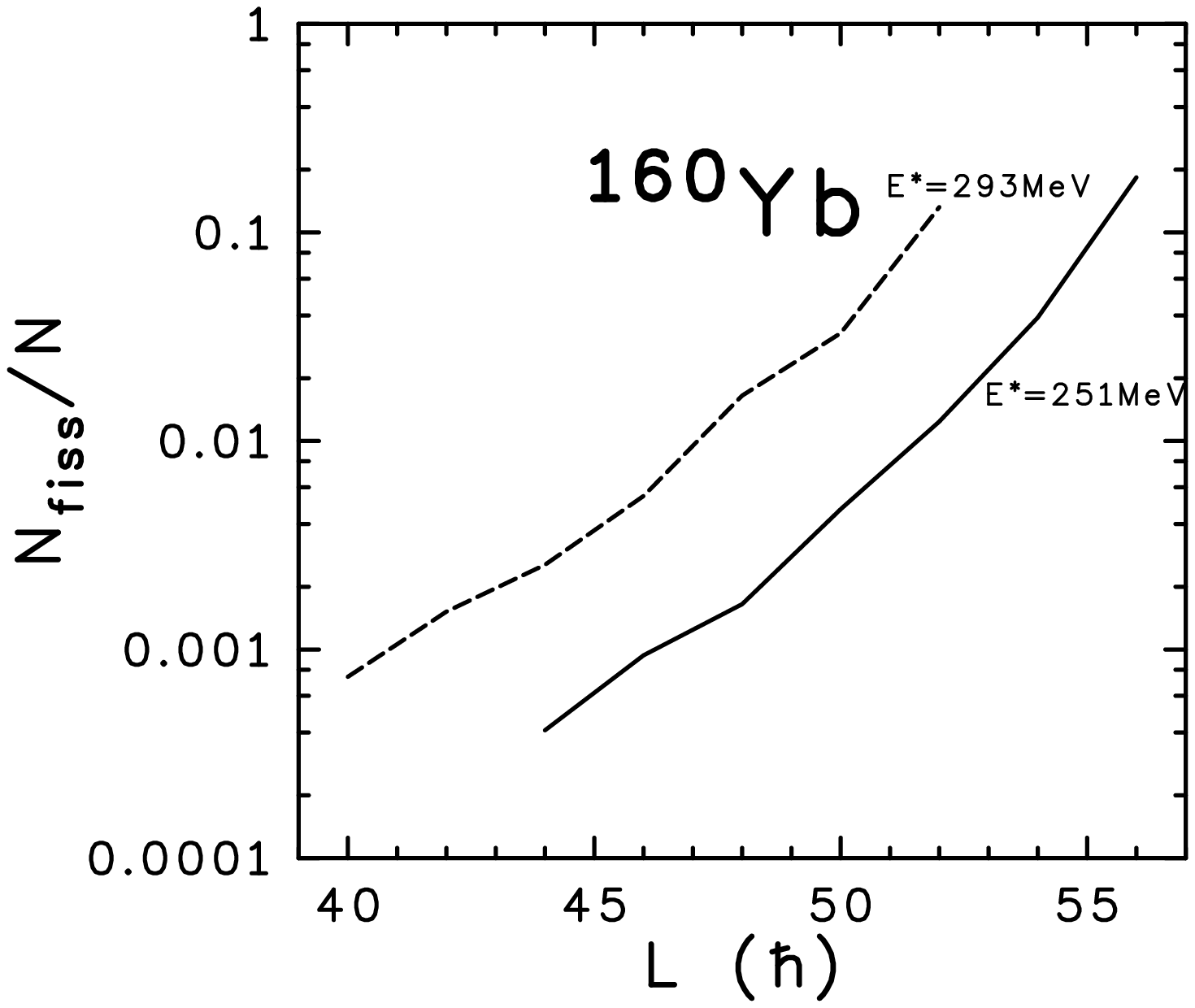}
\end{figure} 
\begin{center}\vspace*{4cm} Figure 24 \end{center}
\pagebreak[4]

\begin{figure}[t]\vspace*{-80mm} \hspace*{20mm}
\epsfxsize=150mm \epsfysize=200mm \epsfbox{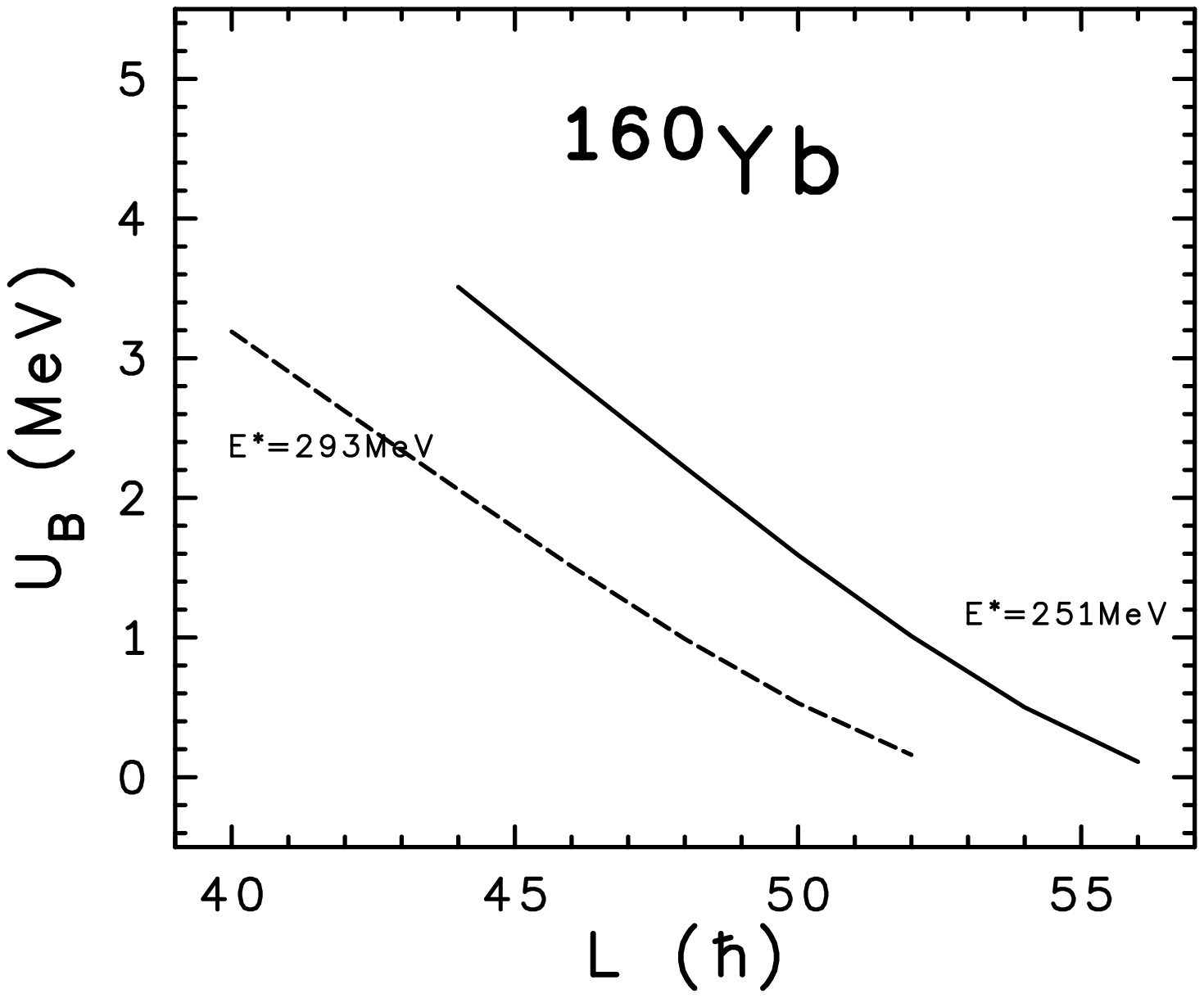}
\end{figure} 
\begin{center}\vspace*{4cm} Figure 25 \end{center}
\pagebreak[4]

\begin{figure}[t]\vspace*{-80mm} \hspace*{20mm}
\epsfxsize=150mm \epsfysize=200mm \epsfbox{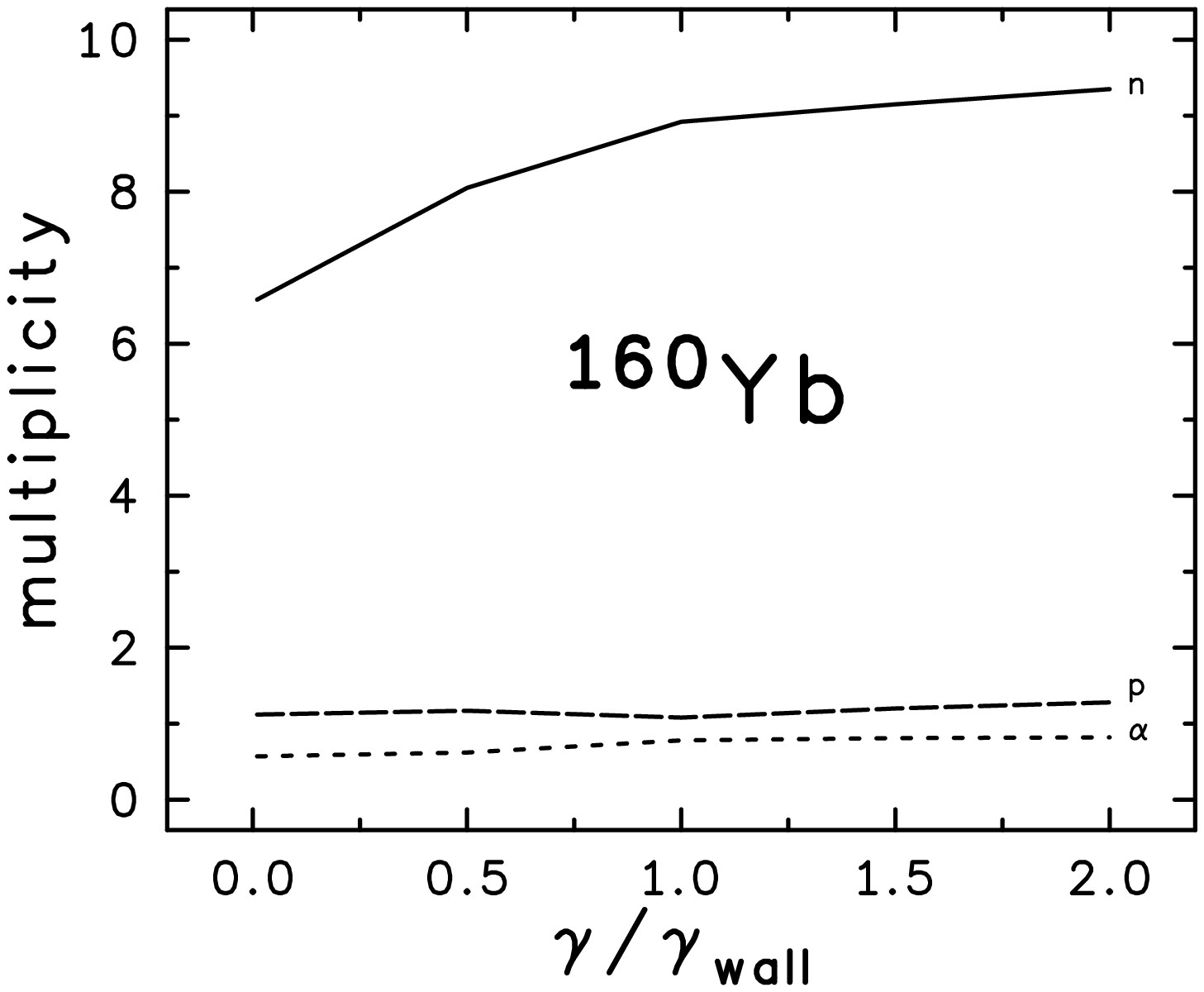}
\end{figure} 
\begin{center}\vspace*{4cm} Figure 26 \end{center}
\pagebreak[4]

\end{document}